\documentclass[]{metastyle}
\usepackage[utf8]{inputenc} %
\usepackage[T1]{fontenc}    %
\usepackage{url}            %
\usepackage{booktabs}       %
\usepackage{amsfonts}       %
\usepackage{nicefrac}       %
\usepackage{microtype}      %
\usepackage{xcolor}         %
\usepackage{soul}

\usepackage{amsmath,amsfonts,bm}

\def\eqref#1{equation~\ref{#1}}

\def\1{\bm{1}}

\DeclareMathAlphabet{\mathsfit}{\encodingdefault}{\sfdefault}{m}{sl}
\SetMathAlphabet{\mathsfit}{bold}{\encodingdefault}{\sfdefault}{bx}{n}

\newcommand{\KL}{D_{\mathrm{KL}}}
\newcommand{\JS}{D_{\mathrm{JS}}}

\newcommand{\Cov}{\mathrm{Cov}}

\newcommand{\kcalm}{kcal\,mol$^{-1}$\xspace}

\DeclareMathOperator{\Tr}{Tr}

\usepackage{bm}
\usepackage{parskip}
\usepackage{dsfont}
\usepackage{tikz}
\usepackage{amsmath}
\usepackage{amssymb}
\usepackage{makecell}
\usepackage[version=4]{mhchem}
\usepackage{xr}
\usepackage{algorithm}
\usepackage[noend]{algpseudocode}
\usepackage{algorithmicx}
\usepackage{xr-hyper}
\usepackage{hyperref}
\usepackage{multirow}
\usepackage{placeins}
\usepackage{xspace}
\usepackage{import}

\newcommand{\dg}{$\Delta G_{\mathrm{W} \mapsto \mathrm{O}}$\xspace}
\newcommand{\dgaa}{$\Delta G^{\mathrm{AA}}_{\mathrm{W} \mapsto \mathrm{O}}$\xspace}
\newcommand{\dgcg}{$\Delta G^{\mathrm{CG}}_{\mathrm{W} \mapsto \mathrm{O}}$\xspace}
\newcommand{\jun}{\textit{juniper}\xspace}

\newcommand{\mar}{MARTINI\xspace}
\newcommand{\alg}{ALOGPS\xspace}
\newcommand{\mmd}{$\mathrm{MMD}^{2}$\xspace}

\renewcommand{\cite}{\citep}

\title{Recovering molecules from coarse-grained beads: free-energy-conditioned generative backmapping across chemical space}

\author[1]{Luis Itza Vazquez-Salazar}
\author[1,2]{Tristan Bereau}

\affiliation[1]{Institute for Theoretical Physics, Heidelberg University, Heidelberg, Germany.}
\affiliation[2]{Interdisciplinary Center for Scientific Computing (IWR), Heidelberg University, Heidelberg, Germany.}

\abstract{
Transferable coarse-grained (CG) force fields compress chemical space: by aggregating atoms into a reduced set of interaction beads, models such as \mar reduce the number of distinguishable compounds by roughly three orders of magnitude, making high-throughput screening of thermodynamic properties tractable across soft matter, with drug--membrane permeability as a well-developed example. The compression is lossy and, so far, one-way: a screen returns a combination of beads, with no established route back to the compounds it stands for. Recovering those compounds---\emph{compositional} backmapping---is a one-to-many inverse map, distinct from the better-studied conformational problem of rebuilding atomic coordinates from a known mapping. Here we formulate compositional backmapping as conditional graph generation by introducing \jun, a discrete denoising diffusion model over molecular graphs conditioned on the octanol--water partition free energy \dg, the principal driver of \mar bead type assignment and hence a proxy for bead identity. Trained on molecules of up to 9 heavy atoms mapped onto one or two beads, \jun generates molecules that are 93\% valid and 92\% unique for two-bead targets, and whose \dg distributions track the target \dgcg linearly ($r^{2} \geq 0.96$), departing only in the hydrophobic and hydrophilic tails. Although the model receives no chemical information beyond a single scalar, the functional groups shift systematically with the imposed free energy, from branched hydrocarbons at the apolar end to amides, imides, and isocyanates at the polar end. A bead combination flagged by a CG screen can therefore be turned into candidate molecules for atomistic study or synthesis.
}
\correspondence{LIVS (\email{l.i.vazquez-salazar@thphys.uni-heidelberg.de})}

\begin{document}
\makeatletter
\providecommand{\bibdata}[1]{}
\providecommand{\bibstyle}[1]{}
\let\xr@save@bibcite\bibcite
\let\xr@save@bibstyle\bibstyle
\let\xr@save@bibdata\bibdata
\renewcommand{\bibcite}[2]{}
\renewcommand{\bibstyle}[1]{}
\renewcommand{\bibdata}[1]{}
\makeatother
\makeatletter
\let\bibcite\xr@save@bibcite
\let\bibstyle\xr@save@bibstyle
\let\bibdata\xr@save@bibdata
\makeatother

\maketitle

\section{Introduction}

Chemical compound space is the set of all possible molecules or materials~\cite{coley2021defining}. Given its definition, chemical space is vast: after applying physical constraints and restricting the possible composition to the elements C, N, O, P, S, F, Cl, Br, and I with a molecular weight of less than 1000\,Da, its size has been hypothesised to be around $10^{200}$~\cite{restrepo2022chemical,gorse2006diversity}. Even after considering that only one in $10^{20}$ compounds is stable, the size of this ``constrained'' chemical space is reduced to $10^{180}$, which is still larger than the estimated total amount of information in the visible universe ($10^{123}$)~\cite{vopson2021estimation}. If chemical space exploration continues at the present rate, discovering the compounds of even this constrained set would take on the order of 10,000 years~\cite{restrepo2022chemical}. Enumeration is therefore not a strategy, and the practical question becomes one of search: how to impose a target property and recover the compounds that satisfy it.

\begin{figure}
    \centering
    \includegraphics[width=\textwidth]{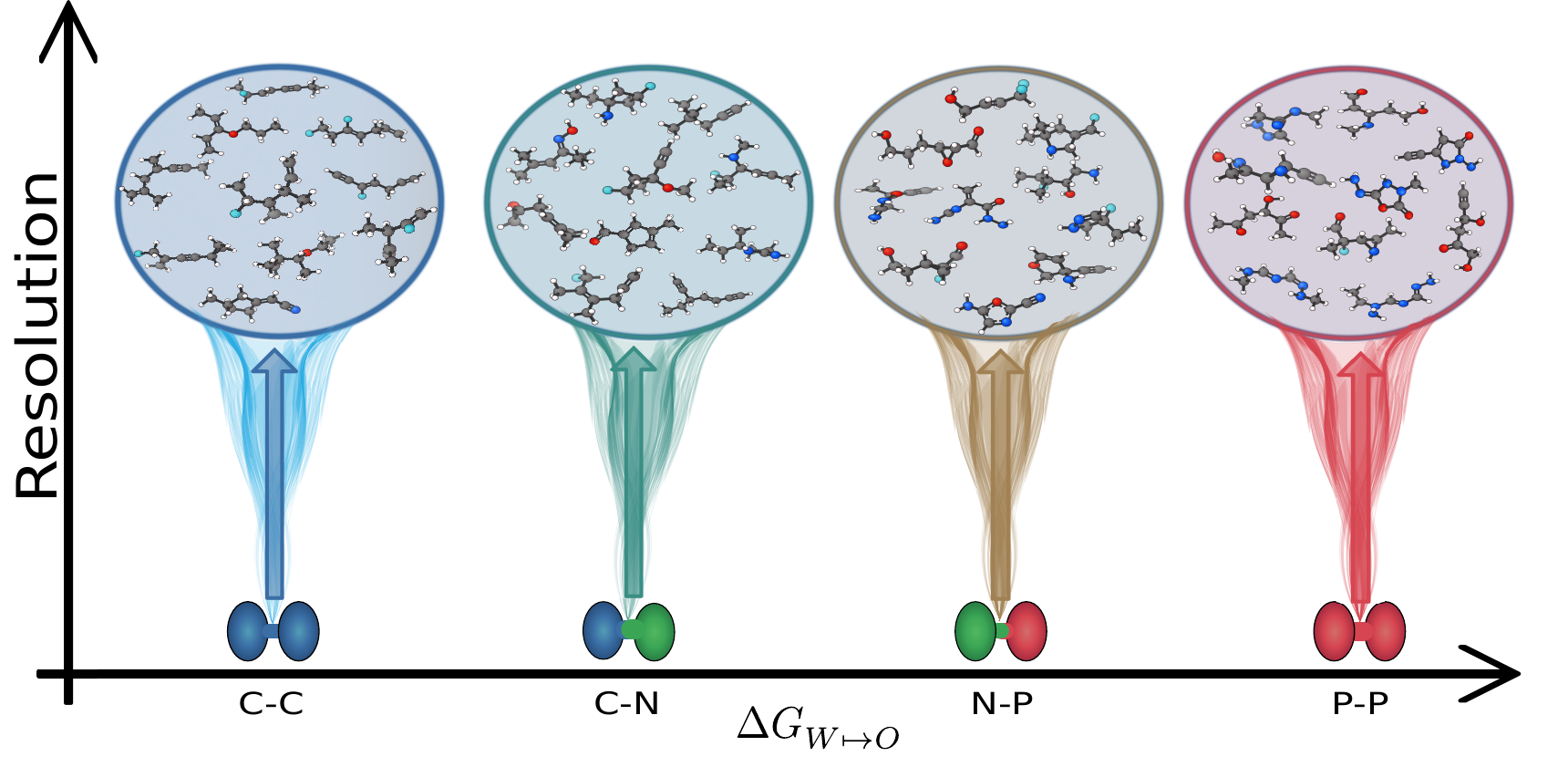}
    \caption{\textbf{Backmapping in chemical space.} Four combinations of the main set of MARTINI beads are shown. The C bead (blue) is apolar, N (green) is intermediate, and P (red) is polar. The horizontal axis orders the combinations by the octanol--water partition free energy (\dg), from the apolar C-C pair on the left to the polar P-P pair on the right; the vertical axis indicates resolution, with the bead pair at the bottom of each panel and the corresponding molecules above. Because each bead type comprises several subtypes, the four combinations are representatives placed schematically along the \dg axis rather than at their nominal values. The backmapping procedure converts each bead combination into a set of molecules with defined physicochemical properties, through a generative process constrained by the value of \dg. Molecular examples are selected at random for each combination.}
    \label{fig:bkmap}
\end{figure}

An alternative for exploring this large space is the use of coarse-grained (CG) models, which lower the dimensionality of the problem by aggregating atomistic degrees of freedom (DOF) into interaction centres called beads~\cite{ingolfsson2014power}. For transferable CG models---such as MARTINI~\cite{marrink2007martini} and SIRAH~\cite{klein2023sirah}---the reduction of DOF produces a reduction in the size of chemical space (Figure~\ref{fig:bkmap}). This reduction greatly aids the exploration of chemical space~\cite{bereau2021computational,menichetti2019drug,hoffmann2019controlled}: a decrease by roughly three orders of magnitude makes studies involving high-throughput screening considerably faster, since one combination of beads represents multiple chemical compounds, which in practice translates to the testing of multiple compounds at once. The same compression underpins molecular design, since a search space small enough to be traversed iteratively admits active learning over CG chemical space~\cite{mohr2022data} and Bayesian optimisation across levels of resolution~\cite{walter2025navigating}, both of which propose bead combinations that optimise a target property instead of ranking an enumeration fixed in advance. The reduction is also lossy. A bead combination stands for a whole family of molecules, and the map from compounds to beads discards the chemical detail that a chemist needs in order to act on a screening or optimisation result.

Beyond this compression, popular CG force fields enable the simulation of a wide range of systems in the biomolecular and materials sciences~\cite{noid2023persp,marrink2013perspective}. Among the existing force fields, MARTINI has established itself as one of the most widely used, because of its flexibility and its large range of applications in biomolecular processes~\cite{souza2020protein,souza2021perspectives,bartocci2024millisecond,stevens2023molecular} and materials science~\cite{vazquez2020martini,alessandri2021martini}. MARTINI uses a mixture of top-down and bottom-up parametrisation, tuning the Lennard-Jones parameters to match experimental partition coefficients and fitting the bonded potentials to all-atom (AA) simulations~\cite{marrink2013perspective,souza2021martini,marrink2023two}. One reason for the popularity of MARTINI is that generating a new parametrised molecule is relatively straightforward, because of its simple construction principles.

However, the inverse process, called ``backmapping,'' in which the molecules corresponding to a certain number of beads are obtained, is far more difficult (Figure~\ref{fig:bkmap}). This difficulty has two main sources. Let $\mathbf{M}$ denote the coarse-graining operator, so that the $\mathbf{M}^{-1}$ operator
reconstructs an all-atom (AA) configuration from a CG one. (i) A one-to-many transformation is in general necessary, $\mathbf{M}^{-1}:\mathbb{R}^{\mathrm{beads}} \mapsto \mathbb{R}^{\mathrm{atoms}}$ with $\dim(\mathrm{atoms}) > \dim(\mathrm{beads})$, so that the dimension of the AA representation always exceeds that of the CG one, implying that certain information needs to be guessed. (ii) In transferable CG models---and in chemical space---the transformation is non-injective and probabilistic, $\mathbf{M}^{-1}: \mathbb{R}^{\mathrm{beads}} \rightrightarrows \mathbb{R}^{\mathrm{atoms}}$, which implies that each CG representation corresponds to multiple AA molecules.

Two distinct problems travel under the single term ``backmapping.'' \emph{Conformational} backmapping reconstructs atomic coordinates from a CG configuration whose mapping $\mathbf{M}$ is already known; it has absorbed most of the methodological effort and is by now well developed~\cite{jones2025flowback,hummerich2025split}. \emph{Compositional} backmapping asks the prior question---which molecules correspond to a given combination of beads---and has remained largely unexplored: the non-injectivity of point (ii) has been sidestepped rather than solved. A high-throughput screen or a CG optimisation loop returns a bead combination, and without the compositional map such a combination cannot be turned into a chemical compound to study atomistically or to synthesise. Compositional backmapping therefore closes the multiscale discovery loop: after chemical space is compressed and screened at low resolution, it recovers the atomistic molecules corresponding to the optimal CG candidates. This work addresses such a compositional problem. 

In this work, we propose to tackle the problem of backmapping in chemical space by using methodologies of generative machine learning (ML). These models aim to learn the underlying statistical distribution of a set of data to later generate new samples that resemble those in the training dataset~\cite{du2024machine,tomczak2024hybrid}. By construction, generative ML offers an elegant solution for the backmapping problem in chemical space because, by learning the underlying data distribution, it is possible to obtain a general map between beads and atoms from a subset of curated samples. Furthermore, because the model generates a distribution that resembles the training dataset, it allows us to generate multiple compounds starting from a single label, effectively bypassing the one-to-many problem. Previous work has successfully used generative ML models to backmap spatial coordinates from CG to AA, both for condensed-phase molecular liquids and polymer melts~\cite{stieffenhofer2020adversarial,li2020backmapping,stieffenhofer2021adversarial} and, predominantly, for peptides and proteins~\cite{wang2022generative,shmilovich2022temporally,arts2023two,jones2023diamondback,jones2025flowback,ugartelatorre2025cgback,berlaga2025flowbackadjoint,hummerich2025split}.

The MARTINI model parametrisation is mainly driven by two aspects: hydrophobicity and preserving the volume, shape, and symmetry of the underlying AA representation. By considering these principles, in this work, we focus on the backmapping of small molecules with up to 9 heavy atoms that are mapped onto one or two \mar beads, according to \mar guidelines of
a maximum of 4 heavy atoms per bead and a maximum mismatch of 1 heavy atom. The setting is deliberately controlled: it isolates the compositional problem from the conformational one, leaving chemistry as the only thing the model must learn. The restrictions carry two consequences. First, the admissible \mar representations for the considered molecules are reduced to only two possible geometries: a point particle (one bead) or a straight line (two beads). This reduction diminishes the importance of the CG spatial representation and allows us to formulate the problem in molecular graph space rather than as a 3D coordinate problem. The 3D representation nevertheless has a large impact on the use of the \mar model~\cite{alessandri2019pitfalls}; the restriction was imposed to simplify the problem of backmapping across chemical space. The second consequence, and the most important for the present study, is that the octanol--water partition free energy \dg (the hydrophobicity descriptor) emerges as the principal parameter controlling how MARTINI compresses chemical space. We accordingly adopt \dgcg both as a constraint for molecular generation and as a proxy for the MARTINI CG mapping. However, \dg has limitations in separating chemical space~\cite{kanekal2019resolution}, and degeneracies (i.e. multiple bead combinations can be mapped to the same value of \dgcg) are expected. A complementary reason for choosing \dg as a driver for our backmapping procedure is the large number of studies that use \dg or $\log P$ (the logarithm of the octanol--water partition coefficient) for molecule generation and optimisation~\cite{gomez2018automatic,lim2018molecular,kang2019conditional,kotsias2020direct,jin2018junction} given the importance of hydrophobicity in drug design and discovery~\cite{arnott2012influence,lobo2020there}.

Many of the existing generative models conditioned on \dg are based on variational autoencoders (VAEs)~\cite{kingma2019introduction}, a generative algorithm that maps a variable to a low-dimensional space and later reconstructs the original variable from it. During training, the model learns the reduction and reconstruction processes. Because of their simplicity, VAEs are highly flexible; however, this flexibility comes at the cost of problems such as posterior collapse, difficulty capturing complex distributions, or incomplete reconstructions~\cite{tomczak2024hybrid}. As an alternative to VAEs, diffusion-based models have emerged~\cite{sohl2015deep,song2020score,ho2020denoising,kingma2021variational}. Diffusion models follow the same strategy as VAEs, except that the forward process (transformation to a simple latent space) is dictated by non-equilibrium statistical mechanics, while the reconstruction process is also learned, making training more stable, providing better reconstructions, and enabling easier conditioning. A few models conditioning diffusion on $\log P$ can be found in the literature~\cite{oestreich2025drugdiff,zhang2025mg,nisonoff2025unlocking}.

It should be noted that the objective of this work differs from the mentioned previous uses of \dg-conditional generation in the literature, which were oriented toward optimising a chemical structure with respect to an arbitrary target property. Our goal is fundamentally different: to recover a distribution of molecules consistent with a \mar representation, in both \dg and chemical diversity. The conditioning value \dgcg is therefore treated as (approximately) the mean of that distribution, rather than as a target to be hit by an individual molecule. Because we are concerned here with small molecules represented as chemical graphs, a graph-based diffusion model whose generation is constrained by \dgcg is a natural choice. The model used here for such a purpose is called \jun (see the SI for an explanation of the name).

In the rest of the text, we present the details of the constructed model and the training procedure. Furthermore, we describe how chemical space was sampled based on the value of \dgcg, which ultimately corresponds to MARTINI beads. We analyse the generated results in terms of physicochemical properties, such as the distribution of \dg, and the chemical space explored. In Section \ref{sec:results}, the unconditional generation results are discussed first, followed by the conditional generation of molecules. Finally, we draw conclusions and outline avenues for further research.

\section{Methods}

\subsection{Diffusion models in graphs}
\label{subsec:diffusion}
For the purpose of this work, we consider molecules at the AA level as graphs $G$ composed of nodes and edges, $G = (\mathcal{X},\mathcal{E})$, where $\mathcal{X}$ denotes the nodes (atom types) and $\mathcal{E}$ the corresponding edges (bond types). Furthermore, to each graph a global label $y$ is assigned, here the hydrophobicity represented by the scalar value \dgcg, so that each molecule is represented as $G = (\mathcal{X},\mathcal{E},y)$. In the rest of the text, the terms ``molecule'' and ``graph'' are used as synonyms.

The diffusion model used in this work operates on the discrete space created by the nodes and edges of the molecules. Because of the discrete nature of the problem, typical diffusion methods, such as denoising diffusion probabilistic models~\cite{ho2020denoising}, are not directly applicable. Instead, here the discrete denoising diffusion methodology introduced by Vignac \textit{et al.}~\cite{vignac2022digress} in the \textsc{DiGress} model is used.

\begin{figure}[h]
    \centering
    \includegraphics[width=\textwidth]{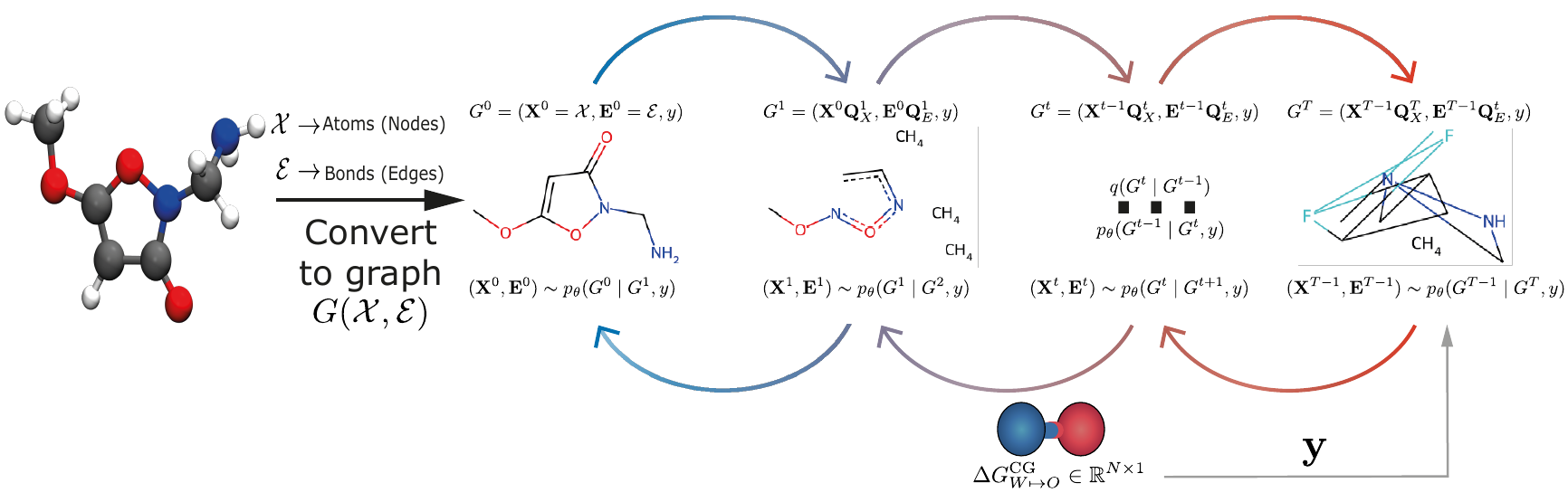}
    \caption{\textbf{Diffusion model for backmapping.} An initial molecule is transformed into a graph $G = (\mathcal{X},\mathcal{E})$. The obtained graph is passed through the forward process (left to right), in which noise is progressively added to nodes and edges over $T$ steps, updating the marginal distributions of nodes and edges in the training dataset. The inverse process involves obtaining denoising probabilities from a transformer model conditioned on \dgcg.}
    \label{fig:diffusion}
\end{figure}

Training proceeds in two stages, summarised in Figure~\ref{fig:diffusion}. In the forward process, noise is added to the nodes and edges of the input graph over $T$ steps, with transition matrices constructed so that the corrupted graph converges to the marginal distributions of atom and bond types in the training dataset rather than to a uniform distribution. In the reverse process, a graph transformer~\cite{dwivedi2020generalization} $\Phi_{\theta}$ is trained to iteratively recover the clean graph from a noisy one, conditioned on the global label $y$. The network uses 5 layers with 8 attention heads and input channels of size 256, 128, and 128 for $\mathbf{X}$, $\mathbf{E}$, and $y$, respectively. There are different ways of computing the denoising probabilities; here, the classifier-free procedure introduced by Ninniri \textit{et al.}~\cite{ninniri2024classifier} was used, which rests on the separation of conditional probabilities formulated by Dhariwal and Nichol~\cite{dhariwal2021diffusion}. The full formulation is given in the SI, Section ``\nameref{sisec:diffusion}''.

The strength of the conditioning enters through a single hyperparameter, which the Results examine in detail. Following Ho and Salimans~\cite{ho2022classifier}, the network output is expressed as a barycentric combination of conditional and unconditional predictions,

\begin{equation}
    \hat{P}_{\theta}(G^{t-1} \mid G^{t},y) = (1-\gamma)\cdot\underbrace{p_{\theta}(G^{0} \mid G^{t},y_{\varnothing})}_{\text{unconditional}} {}+{} \gamma\cdot\underbrace{p_{\theta}(G^{0} \mid G^{t},y)}_{\text{conditional}}.
    \label{eq:cond}
\end{equation}

Here, $\gamma \geq 0$ is the guidance strength. To make the unconditional branch available and prevent the model from ignoring the conditioning~\cite{ho2022classifier,ninniri2024classifier}, the conditioning value $y$ is replaced during training by a learnable null value $y_{\varnothing}$ with probability $\rho$, set here to 0.2.

\subsection{Training}

The model was trained using the unimer (one bead) and dimer (two beads) datasets obtained from Ref.~\citenum{menichetti2019drug}, consisting of molecules with up to 9 heavy atoms (C, N, O, F) mapped to 28 beads of the MARTINI CG force field v2~\cite{marrink2007martini}, corresponding to the polar (P), intermediate (N), and apolar (C) blocks in normal and small sizes. Within the apolar and polar blocks, a numerical subtype running from 1 to 5 denotes increasing polarity, so that C1 is the most apolar bead of the set and P5 the most polar. The intermediate block is instead labelled by hydrogen-bonding character: acceptor (Na), donor (Nd), both (Nda), or neither (N0). A pair such as C1-C1 therefore denotes the most hydrophobic two-bead representation available, and P5-P5 the most hydrophilic. The datasets were generated from the GDB database~\cite{ruddigkeit2012enumeration} using \textsc{Auto-Martini}~\cite{bereau2015automated}. The values of \dg used as global labels correspond to those obtained for the individual CG beads and to their sum in the multiple-bead case.

The training database was curated to meet specific thresholds for the CG molecular representation. For single-bead representations, each bead contained at most 5 heavy atoms. For two-bead representations, only molecules with 6 to 9 heavy atoms were considered. Given the distribution of chemical space and the inherent biases in the parent database of our training set~\cite{glavatskikh2019dataset,vazquezsalazar2021}, an imbalance among bead representations was expected (see Figure~\ref{sifig:train_hist}).

During the training process, the following loss function was optimised:

\begin{equation}
\mathcal{L}(\mathbf{X},\mathbf{E},y) = \mathrm{CE}(\mathbf{X}^{0},\hat{P}_{\theta}(\mathbf{X}^{t})) + \lambda\cdot \mathrm{CE}(\mathbf{E}^{0},\hat{P}_{\theta}(\mathbf{E}^{t})) + \eta \cdot \mathrm{MSE}(y^{0},\hat{P}_{\theta}(y)),
\end{equation}

where $\mathrm{CE}$ denotes the cross-entropy loss, $\mathrm{MSE}$ the mean squared error, and $\lambda$ and $\eta$ are hyperparameters that weight the influence of the different parts of the graph, set to $\lambda=5$ and $\eta=1$. The training process was run for 1000 epochs, each with $T=500$ diffusion steps, with a cosine diffusion schedule. The optimiser used was AdamW~\cite{loshchilov2017fixing} with a learning rate of $10^{-4}$ and a weight decay of $10^{-9}$.

\subsection{Generation}

The obtained model was used to generate molecules with a given value of \dgcg starting from random noise. The denoising procedure is also conditioned on the number of nodes in the graph (i.e., the number of atoms), which is drawn from the node distribution in the training dataset for either unimers or dimers. As a baseline, 1000 samples were generated without guidance.

Unless otherwise stated, 100 molecules were generated for each bead combination present in the training dataset. In total, 1,200 samples were generated for unimers and 38,600 samples for dimers. The analysis that follows concentrates on dimers; the unimer results are limited by their much smaller population, though the behaviour observed for dimers is expected to carry over to them. For the analysis, the labels were treated as permutation-invariant (i.e., A-B and B-A were considered equivalent) and the bead size was ignored. This process was repeated for multiple values of $\gamma \in \lbrace 0.5, 1, 2 \rbrace$ in Eq.~\ref{eq:cond}. The values of $\gamma$ were picked to illustrate the different regimes of generation. In the case of $\gamma = 1$, the model directly samples the conditional distribution, $\hat{P}_{\theta}(G^{t-1} \mid G^{t},y)$. For $\gamma = 0.5$, there is an equal contribution from the conditional and unconditional probabilities. Finally, for $\gamma = 2$, the weight of the conditional distribution is doubled, and the unconditional distribution works as a correction to the former. Ho and Salimans report improved sample quality for $\gamma > 1$~\cite{ho2022classifier}.

\subsection{Analysis}

\subsubsection{Basic metrics}
\label{subsec:basic_met}
To evaluate the performance of a generative model, we use a set of standard metrics. In this work, we consider a generated molecule valid if all its fragments are connected and it can be reconstructed from its graph representation by RDKit~\cite{landrum2013rdkit}. From the valid molecules, we measure the number of unique generated molecules as $N_{\mathrm{unique}} = |\lbrace \mathrm{Mols}_{\mathrm{valid}}\rbrace|$, the number of distinct valid molecules after removing duplicates, where $|\cdot|$ denotes cardinality (i.e., the number of unique elements in a set). Similarly, the number of novel molecules is obtained as $N_{\mathrm{novel}} = |\mathrm{Mols}_{\mathrm{unique}} \setminus \mathrm{Mols}_{\mathrm{train}}|$, where $\setminus$ represents the set difference (i.e., the elements of one set that do not belong to the other).

Based on the number of valid, unique, and novel molecules, it is possible to define quantities that characterise the performance of the generative model in chemical space. Here, the following quantities were studied~\cite{zhang2021comparative}:
\begin{itemize}
    \item \textit{Repetition rate}: this metric quantifies the number of duplicates generated by the model. It is defined as
    \begin{equation}
        R_{\mathrm{repeat}} = \frac{N_{\mathrm{valid}}-N_{\mathrm{unique}}}{N_{\mathrm{unique}}}.
        \label{eq:repeat}
    \end{equation}

    \item \textit{Coverage}: this quantity measures how much of the chemical space of the training dataset is reproduced by the generative model. It is defined as
    \begin{equation}
        \Cov = \frac{N_{\mathrm{unique}}-N_{\mathrm{novel}}}{N_{\mathrm{dataset}}}.
        \label{eq:coverage}
    \end{equation}

    where $N_{\mathrm{dataset}}$ is the total number of samples in the training dataset. 
    
    \item \textit{Novelty rate}: this quantity measures the proportion of molecules produced by the generative model that were absent from its training set. It is defined as
    \begin{equation}
        R_{\mathrm{novel}} = \frac{N_{\mathrm{novel}}}{N_{\mathrm{unique}}}.
        \label{eq:novel}
    \end{equation}

\end{itemize}

\subsubsection{Distribution analysis}
\label{subsec:dist_ana}
The generated samples were analysed to evaluate the physicochemical characteristics of the molecules and their chemical space coverage. The molecules were generated using the scalar partition free energy between octanol and water (\dgcg), one of the main drivers of MARTINI model parametrisation. This property therefore serves to evaluate our procedure. The determination of \dg is challenging; here, two estimators of the octanol--water partition coefficient $\log P$ were used. The first is the Wildman--Crippen method~\cite{wildman1999prediction} as implemented in RDKit~\cite{landrum2013rdkit}, an additive scheme that sums contributions assigned to individual atoms. The second is \alg~\cite{tetko2002application}, an associative neural network whose mapping to $\log P$ is not constrained to be additive, accessed through its web interface. We quote \alg values throughout, on the grounds that partitioning is a collective property of the solute and its solvation shell and is therefore poorly served by an additive atom-contribution scheme; the corresponding RDKit results are collected in the Supporting Information. Regardless of the model, the value of $\log P$ is related to \dg by the following expression:

\begin{equation}
    \Delta G_{\mathrm{W}\mapsto \mathrm{O}} = -RT\ln(10)\log P,
\end{equation}

where $R=1.987\times 10^{-3}$\,kcal\,mol$^{-1}$\,K$^{-1}$ and $T=300$\,K. The values of \dgaa were computed for all generated samples and, in the case of the Wildman--Crippen method, recomputed for the molecules in the training dataset.

The generated and training distributions of \dgaa were compared using the Jensen--Shannon divergence ($\JS$). This quantity is symmetric and measures the total divergence from the mean distribution, as it equals the average divergence of each distribution from the arithmetic mean of the distributions~\cite{nielsen2019jensen}. $\JS$ is defined as

\begin{equation}
\begin{split}
\JS[p_{\mathrm{train}} \parallel q_{\mathrm{gen}}]
&= \tfrac{1}{2}\KL\!\left[p_{\mathrm{train}}\,\middle\|\,\tfrac{p_{\mathrm{train}}+q_{\mathrm{gen}}}{2}\right]
 + \frac{1}{2}\KL\!\left[q_{\mathrm{gen}}\,\middle\|\,\frac{p_{\mathrm{train}}+q_{\mathrm{gen}}}{2}\right] \\[4pt]
&= \frac{1}{2} \int \mathrm{d}x \left[
p_{\mathrm{train}}(x)\ln\!\left(\tfrac{2p_{\mathrm{train}}(x)}{p_{\mathrm{train}}(x)+q_{\mathrm{gen}}(x)}\right)
+ q_{\mathrm{gen}}(x)\ln\!\left(\tfrac{2q_{\mathrm{gen}}(x)}{q_{\mathrm{gen}}(x)+p_{\mathrm{train}}(x)}\right)\right],
\end{split}
\label{eq:js}
\end{equation}

where $\KL$ is the Kullback--Leibler divergence.

The distributions were also compared using the maximum mean discrepancy (MMD)~\cite{gretton2012kernel}. In this approach, the probability distribution is transformed using a kernel function into a reproducing kernel Hilbert space (RKHS). In the RKHS, the difference between distributions is obtained as the difference between the mean embeddings,

\begin{equation}
    \mathrm{MMD}^{2}(p,q) = \|\mu_{p} - \mu_{q}\|_{\mathcal{H}}^{2}.
    \label{eq:mmd}
\end{equation}

Here $\mathcal{H}$ is the RKHS, and the mean embedding of the distributions (also known as kernel embedding) is given by:

\begin{equation*}
    \mu_{p} = \mathbb{E}_{x\sim p}[k(\cdot,x)] = \int \mathrm{d}p(x)\, k(\cdot,x).
\end{equation*}

A similar expression can be derived for $q$. Here, we used an empirical estimate of the MMD with a Gaussian kernel~\cite{gretton2012kernel}. We also obtained the ``witness function'' of the MMD metric; this function shows how the distributions change over the evaluated range of \dg.

\subsubsection{Functional-group and diversity analysis}
\label{subsec:chem_ana}
To analyse the chemical space of the generated molecules, we identified functional groups with Ertl's algorithm~\cite{ertl2017algorithm}, as implemented by Colmenarejo~\cite{colmenarejo2025efgs} in RDKit. The number and types of functional groups are used to compare the chemical space explored by a given bead combination.

Complementary to the calculation of functional groups, the Fr\'echet ChemNet distance (FCD)~\cite{preuer2018frechnet} is a metric inspired by the Fr\'echet inception distance~\cite{heusel2017gans} commonly used to evaluate generative models. The FCD uses the activations of the penultimate layer of the ChemNet model~\cite{mayr2018large}, trained to predict bioactivity using major drug-discovery databases. ChemNet was chosen because its representation encodes both chemical and biological information. The mean and covariance of the activations are then computed, under the assumption that they follow a multidimensional Gaussian distribution. Then, the reference distribution of molecules ($P_{\mathrm{ref}}$) and the one generated by the model ($P_{\mathrm{gen}}$) are compared using the Fr\'echet distance between the resulting multidimensional normal distributions, given by
\begin{equation}
    \mathrm{FCD}(P_{\mathrm{ref}}, P_{\mathrm{gen}}) = \|\mu_{\mathrm{gen}} - \mu_{\mathrm{ref}}\|_{2}^{2} + \Tr\left[\Sigma_{\mathrm{gen}} + \Sigma_{\mathrm{ref}} - 2(\Sigma_{\mathrm{gen}}\Sigma_{\mathrm{ref}})^{1/2}\right],
\end{equation}

where $\mu_{\mathrm{gen/ref}}$ is the mean of the generated or reference distribution, and $\Sigma_{\mathrm{gen/ref}}$ is the corresponding covariance matrix. Because we are assuming multidimensional normal distributions to describe the ChemNet embeddings, the Fr\'echet distance equals the $L^{2}$-Wasserstein distance.

Lastly, the chemical diversity of the generated molecules is evaluated by the internal chemical diversity (ICD) metric defined by Benhenda~\cite{benhenda2017chemgan} as

\begin{equation}
    \mathrm{ICD} = \frac{2}{N_{\mathrm{gen}}^{2}}\sum_{i}\sum_{j\neq i} \left[1 - \mathrm{JT}(\mathrm{mol}_{i},\mathrm{mol}_{j})\right].
\end{equation}

Here, $\mathrm{JT}$ denotes the Jaccard--Tanimoto similarity between all pairs of generated molecules; the factor 2 accounts for the symmetry of the $i,j$ pairs.

\section{Results}
\label{sec:results}
\subsection{Unconditional generation}
\begin{figure}
    \centering
    \includegraphics[width=0.8\linewidth]{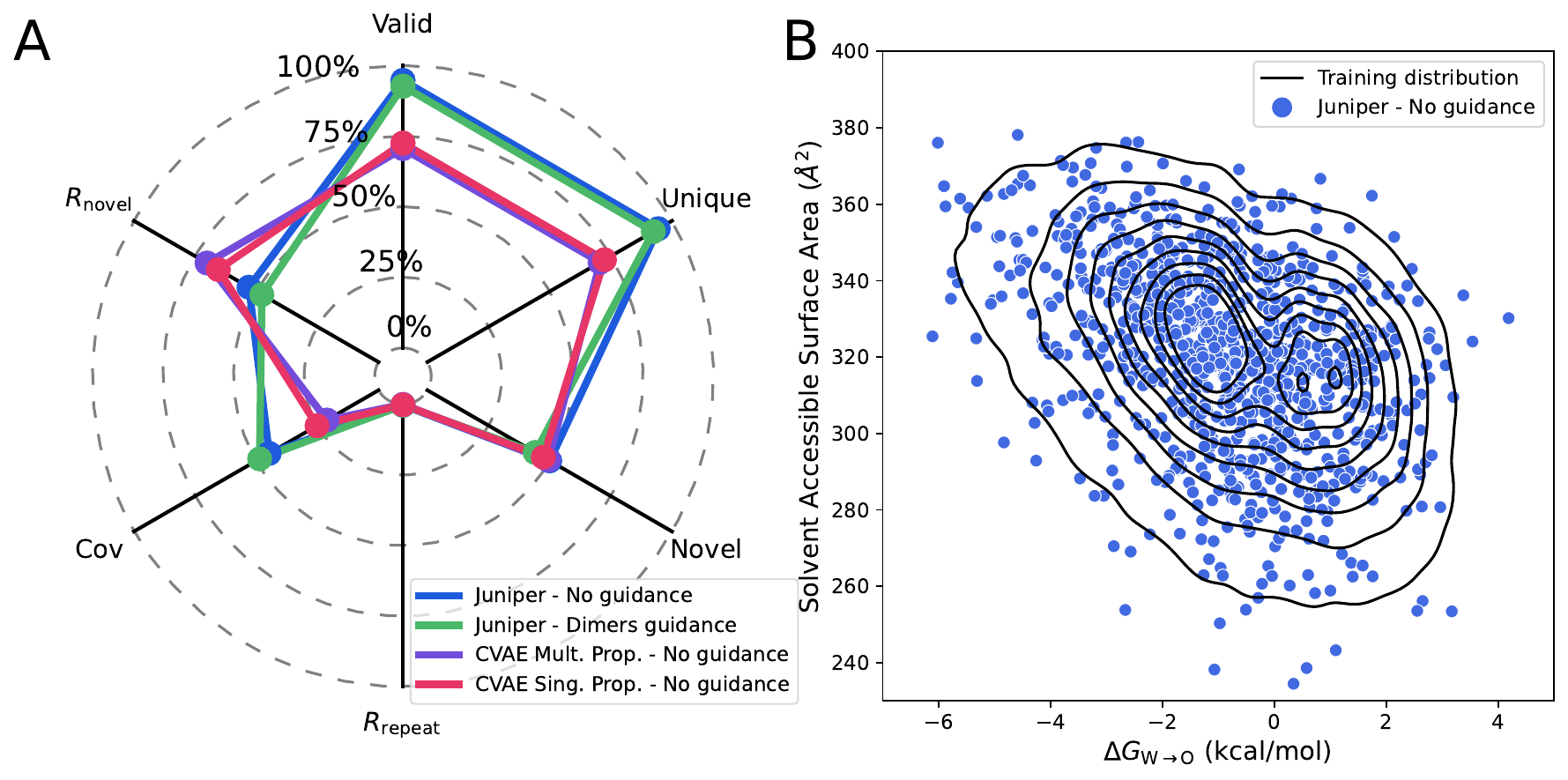}
    \caption{\textbf{Basic metrics and unconditional generation.} (A) Radar chart for the basic performance metrics of the model trained with the coarse-grained dataset. The performance of \jun for generating samples without guidance (blue line). For comparison, we also report the performance of the conditional variational autoencoder (CVAE) in Ref.~\citenum{lim2018molecular} on the same dataset for unconditional generation: models trained with information on multiple properties (purple line) and only with \dg (red line). Finally, the performance for the mean of the generated samples using guidance (green line) for molecules that can be mapped to MARTINI bead dimers is reported. (B) 2D kernel density representation of the chemical space in terms of the octanol--water partition free energy and the solvent-accessible surface area of the training dataset (black line). Samples generated without explicit guidance are shown as blue points. }
    \label{fig:basic}
\end{figure}

We first tested \jun's generative capabilities without explicit guidance, with the aim of observing how the trained model samples the property space defined by \dgcg. To this end, we generated 1000 samples and evaluated a set of basic metrics (Figure~\ref{fig:basic}A). The unconditional \jun model performs well at generating valid and unique compounds, with a low repetition rate. The novelty of the generated molecules is high ($\sim$50\%), albeit at the cost of reduced coverage of the training dataset. The results were benchmarked in two ways. First, the performance was compared with other state-of-the-art models for molecular graph generation, and \jun performs on par with them (Table~\ref{sitable:perfom_other}). This comparison is only indicative, as those models were trained on a different dataset. A second, more direct, comparison was made with the conditional variational autoencoder (CVAE) of Ref.~\citenum{lim2018molecular} (details in the SI). Two CVAE models were trained with our dataset: CVAE Mult. Prop., conditioned on molecular weight, \dgcg, hydrogen-bond donors, hydrogen-bond acceptors, and topological polar surface area; and CVAE Sing. Prop., conditioned on \dgcg alone. Each variant generated 1000 unguided samples to evaluate the same performance metrics.

The comparison in Figure~\ref{fig:basic}A shows that \jun outperforms both CVAE models across all metrics except those related to novelty. In particular, the percentage of valid and unique molecules is $\sim$20\% higher for \jun than for the CVAE models, with the single-property CVAE performing slightly better than the multi-property one. For the reproduction of the training distributions of \dgcg and solvent-accessible surface area, all models yield small Jensen--Shannon divergences ($\JS \sim 10^{-3}$), indicating good agreement between distributions. The values are of comparable magnitude for the two properties. In both cases, \jun attains the lowest value among the models (Table~\ref{sitable:JS_others}). A subtler point is that the \jun distributions resemble those of CVAE Mult. Prop. more than those of CVAE Sing. Prop., despite \jun having been trained on \dgcg alone. Because of the improved performance of \jun with respect to the CVAE models in unconditional generation, only the former is used for the rest of this work. 

Figure~\ref{fig:basic}B projects the chemical space onto the \dg values and the solvent-accessible surface area (SASA). The training distributions of these two properties are drawn as 2D kernel density estimates, and the molecules generated without guidance in blue. The results show that the generated samples uniformly sample the training space, including low-density regions, and reach both high and low \dg, with uniform coverage across the range of SASA values. In comparison, the CVAE models are biased towards the more populated regions; see Figure~\ref{sifig:unguid_samp}. Together, these results show that the model generates molecules of varied shape (characterised by SASA), spanning a range of functional groups and, consequently, different parts of chemical space (as characterised by \dg).

\subsection{Conditional generation with \dg}
\label{subsec:cond_gen}

Having validated the general performance of the trained model, we sampled chemical space at given values of \dg, which correspond to the backmapping of MARTINI CG bead combinations. As for the unconditional generation, we evaluated the basic performance metrics (validity, uniqueness, novelty, repetition rate, and coverage). The results are reported in Figures~\ref{sifig:basic_metrics} and \ref{sifig:rep_nov_rate}. In all cases, the performance of \jun was strong, with 93\% of conditionally generated graphs valid and 92\% unique, as well as low repetition rates; see Figure~\ref{fig:basic}A.

\begin{figure}
    \centering
    \includegraphics[width=\linewidth]{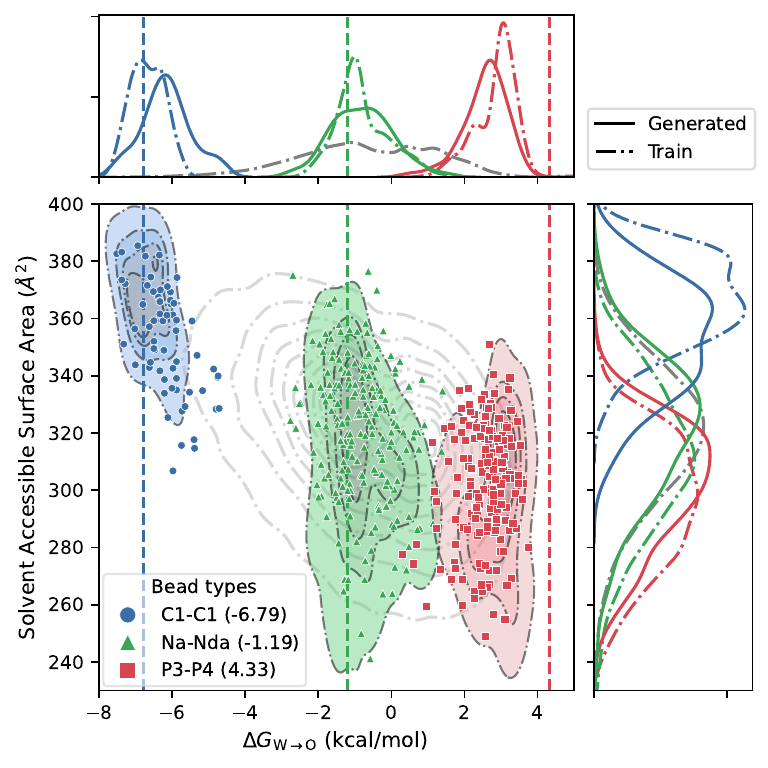}
    \caption{\textbf{Sampling in chemical space.} 2D kernel density estimate of the joint distribution of \dg and the solvent-accessible surface area (SASA). The figure margins show 1D kernel density estimates for each quantity. Solid lines and filled contours refer to generated molecules, dash-dotted lines to the training data. The grey dash-dotted contours in the background show the full training dataset. The filled contours and scattered points represent the conditional samples generated for three cases: C1-C1 (very hydrophobic), Na-Nda (intermediate), and P3-P4 (hydrophilic); the coloured dash-dotted lines in the margins give the training distributions for the same bead combinations, and the vertical dashed lines mark the conditioning value \dgcg of each. The effect of the conditioning on the sampling is visible in all three cases. Values of \dg were obtained with \alg. In the SI, Figure~\ref{sifig:mols_examples} shows examples of the molecules generated for the \dg value corresponding to each bead combination.}
    \label{fig:2d}
\end{figure}

An important aspect of the conditional generation of graphs is that the generated chemical compounds map to specific parts of the chemical space, in contrast to the non-guided sampling that uniformly samples the training distribution, as seen in Figure~\ref{fig:basic}B. Using the low-dimensional representation of chemical space created by \dg and SASA, Figure~\ref{fig:2d} illustrates, for $\gamma=1$, the constrained generation of molecules at the \dg values of three limiting bead combinations. The selected beads were C1-C1, very hydrophobic with a large negative value, \dg $= -6.79$\,\kcalm; Na-Nda, an intermediate value ($-1.19$\,\kcalm) near the overall mean of the distribution; and P3-P4, very hydrophilic with a positive value, \dg $= 4.33$\,\kcalm.

\begin{figure}
    \centering
    \includegraphics[width=0.7\linewidth]{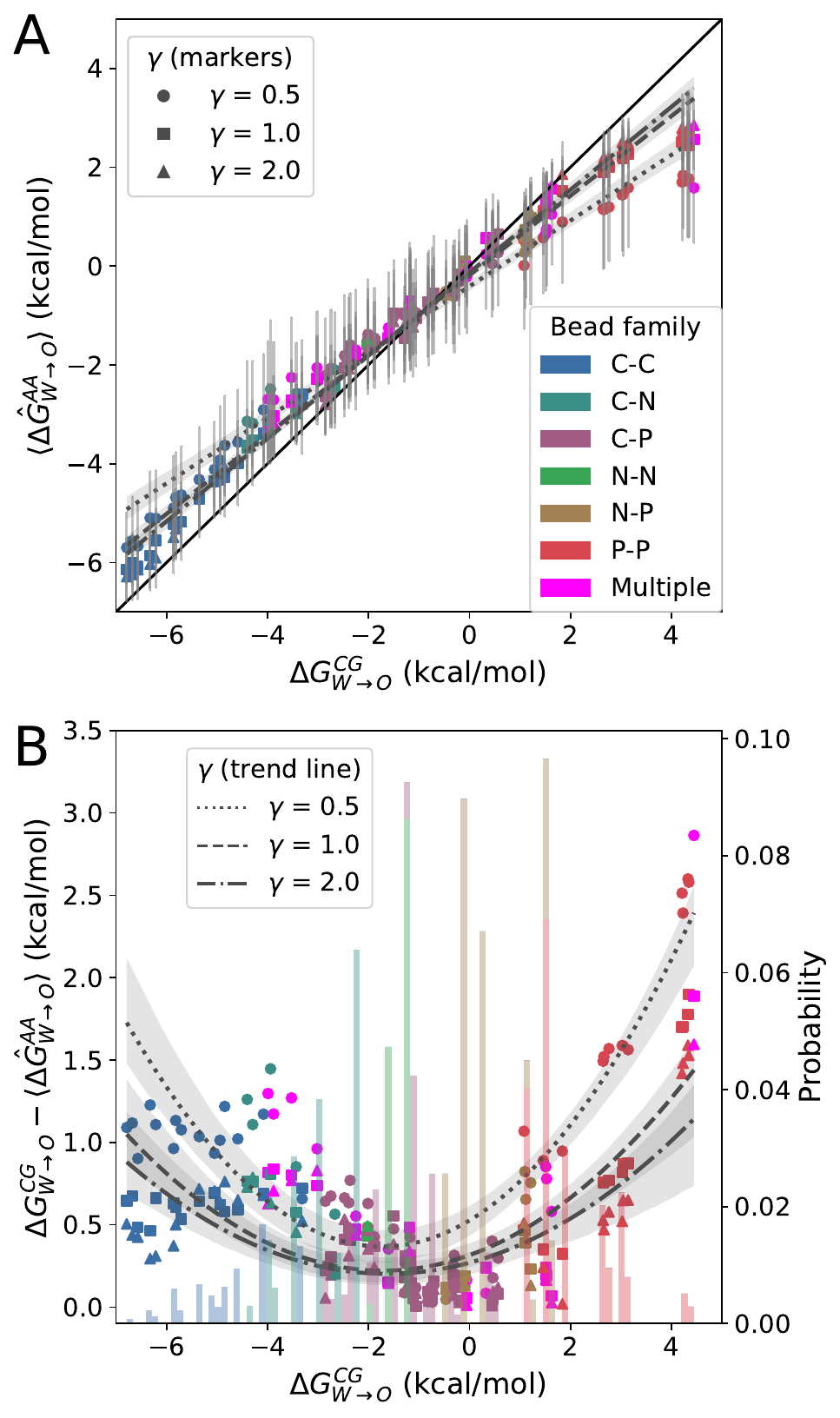}
    \caption{\textbf{Comparison between \dgcg and \dgaa.} (A) Scatter plot of the value of \dgcg versus the mean value of \dgaa for the generated molecules. The markers are coloured by bead family, except for values of \dg that can correspond to multiple bead combinations, which are coloured in magenta. (B) Residual plot of the absolute difference between the mean of \dgaa and the value of \dgcg for all combinations; the signed version is available in the SI. A trend line fitted to a second-order polynomial is shown for each value of $\gamma$. The background shows the histogram of the training dataset.}
    \label{fig:diff_dg_aacg}
\end{figure}

The generated samples illustrate how the model selects specific regions of chemical space characterised by \dg and SASA values. Starting with C1-C1, the mean of the \dg distribution shifts toward negative values, indicating that the generated samples are hydrophobic. A corresponding shift towards larger SASA values indicates that the generated molecules have larger accessible surface areas. Both trends can be explained by the presence of saturated hydrocarbons in the generated molecules, which are known for their hydrophobicity and also contribute more apolar surface area. As expected, the model therefore generates large molecules with saturated hydrocarbons to satisfy the hydrophobicity requirement; see Figure~\ref{sifig:mols_examples}A. C1-C1 lies at the extreme hydrophobic tail of the training distribution, where data are sparse. As a consequence, some of the generated samples lie outside the training distribution, showing that the model interpolates from the nearest, more populated regions. The shift of the distribution mean toward the high-population region is further evidence, at the cost of a large difference between \dgcg and the mean \dgaa of the generated samples (see Figure~\ref{fig:diff_dg_aacg}B).

Similar results are observed for the Na-Nda and P3-P4 combinations. For the intermediate pair, the distribution peak is near the mean of the training distribution; however, most generated molecules lie within a narrow range of $-3$ to $2$\,\kcalm, localised at the centre of the \dg distribution. The SASA value for this combination indicates that molecules are generated across the full range of values, as there is no need to emphasise the presence of hydrophobic fragments in these samples. This is visible in the example molecules for the Na-Nda combination (Figure~\ref{sifig:mols_examples}B), which are characterised by a mixture of hydrocarbon fragments, multiple bonds, and heteroatoms (N, O). Finally, the polar case illustrated by the P3-P4 combination shows the mirror image of its apolar counterpart, with a distribution of \dg values shifted to the right of the training distribution mean and the largest fraction of values between $\sim$0.7 and 3\,\kcalm. This case also displays possible overlaps in the \dg values of different bead combinations. This is expected, given the limited ability of \dg to separate chemical space~\cite{kanekal2019resolution}. However, distribution overlap can help generate stable chemical compounds in poorly populated classes. Interestingly, the SASA values for these molecules are smaller than those for the other two sets. This results from the formation of small rings and highly polar fragments that meet the hydrophilicity requirement and have small surface areas (Figure~\ref{sifig:mols_examples}C).

The previous analysis illustrated \jun's ability to generate molecules given a value of \dg, which traces back to the corresponding \mar bead combination. A key aspect of this work is that, for each \dg value corresponding to a MARTINI representation, the model is trained to recognise that value as a global graph label. We can then generate a distribution of molecules around the provided value. In other words, the value of \dgcg should be near the mean of the distribution of \dgaa of molecules with similar chemical structures. In Figure~\ref{fig:diff_dg_aacg}A, the correlation plot between the values of \dgcg and the mean value of \dgaa of the generated samples obtained with \alg is shown for all values of $\gamma$ evaluated in this work. A clear linear relationship between the two values is observed, with $r^{2} \geq 0.96$ for all values of $\gamma$. Furthermore, for values of $\gamma = 1,2$, the slope of the linear fit is near 0.8, indicating that \dgcg is correctly approximated by the mean of the generated molecules. In contrast, systematic deviations are observed, with overestimation of the values of \dg in the hydrophobic region ($\sim -3$\,\kcalm) and a corresponding underestimation for the hydrophilic values ($\sim 2$\,\kcalm); see Figure~\ref{sifig:signed_dgaacg}. Part of this systematic deviation is attributable to the estimator. Repeating the analysis with RDKit gives larger differences in both tails (Figures~\ref{sifig:2d_rdkit} and \ref{sifig:diff_dg_aacg_rdkit}), as expected of an additive scheme: the Wildman--Crippen method accumulates errors for highly hydrophobic compounds~\cite{mannhold2009calculation}, and for highly hydrophilic ones the absence of corrections for intramolecular interactions underestimates the hydrophobicity~\cite{rasmussen2023machines}.

The residual plot in Figure~\ref{fig:diff_dg_aacg}B more clearly reveals the systematic shortcomings of the \dg predictor in extreme cases. Regardless of the value of
$\gamma$ used, the residuals adopt a slightly parabolic shape, with the smallest values of
$\Delta G^{\mathrm{CG}}_{\mathrm{W}\mapsto\mathrm{O}} - \langle \Delta \hat{G}^{\mathrm{AA}}_{\mathrm{W}\mapsto\mathrm{O}} \rangle$ occurring near intermediate values of \dgcg. Despite this systematic error, most bead combinations satisfy our hypothesised relationship between \dgcg and the mean \dgaa, with the central block (C-N and N-N combinations) showing differences below 1\,\kcalm.

A further aspect of the \dg estimator's shortcomings is how they are amplified by the value of $\gamma$. For example, when considering the P-P combinations, the largest deviations are observed for the smallest value of $\gamma$ tested (0.5), with P5-P5 having the largest deviation overall (2.9\,\kcalm).
In this regard, Figure~\ref{fig:diff_dg_aacg} provides
a few insights about the effect of $\gamma$. First, $\gamma$ has a negligible influence for bead combinations near the centre of the distribution, but for combinations in the tails the effect is more pronounced. As expected, a small value of $\gamma$ leads to large differences for the extreme cases and flattens the slope in Figure~\ref{fig:diff_dg_aacg}A. On the contrary, a large value of $\gamma$ helps in sampling extreme values, with a considerable reduction of the difference for the P-P combinations. Returning to the P5-P5 combination, a reduction of 1\,\kcalm in the difference $\Delta G^{\mathrm{CG}}_{\mathrm{W}\mapsto\mathrm{O}} - \langle \Delta \hat{G}^{\mathrm{AA}}_{\mathrm{W}\mapsto\mathrm{O}} \rangle$ is observed when going from $\gamma = 0.5$ to $\gamma = 1$; a further increase in $\gamma$ leads to a moderate reduction of 0.3\,\kcalm when going from $\gamma=1$ to 2. The rest of the bead combinations show a similar pattern: a considerable change when $\gamma$ goes from 0.5 to 1 and a moderate/negligible change when going from 1 to 2. Because the benefit of larger $\gamma$ is largely recovered by $\gamma=1$, we adopt this value for the remainder of our discussion.

Further analysis of the generated distributions of \dgaa was also performed to confirm that the generated samples match not only the average quantity but also the overall distribution of values used as reference. The objective of this analysis was to confirm that \jun captures the underlying data distribution of \dgaa. This can also be understood as the model generating similar molecules for a specific value of \dg and, in consequence, for the corresponding combination of \mar beads. The metrics used for this purpose were $\JS$ and \mmd (see Section~\ref{subsec:dist_ana} for details). Results by value of \dgcg and bead combination are shown in Figure~\ref{sifig:stat_dist}.

First, many bead combinations, particularly those containing beads of the intermediate class (Na, Nd, Nda), share the same \dg value (Figure \ref{fig:diff_dg_aacg}). In those cases, we took the average distance for each combination and highlighted it in magenta in the plot. Values of $\JS$ follow a trend similar to that of the residuals in Figure~\ref{fig:diff_dg_aacg}B. Most combinations have a small $\JS$ value, with a mean of 0.07 and most evaluated samples within one standard deviation (0.07) of this value. Nevertheless, the parabolic trend described above is noticeable for combinations of the P5 bead. In the case of P5 combined with beads of the C category, the value of $\JS$ decreases as C reduces its hydrophobicity; the opposite is observed for the P family, where $\JS$ increases with hydrophilicity. The complementary \mmd values show a picture similar to that of $\JS$, though they resolve changes in the width of the distribution more clearly, which can be verified with the use of the witness function (Figure~\ref{sifig:witness_s1}). In particular, the \mmd values for combinations of C1 and C2 with beads from the intermediate class are larger because of the width of the generated distributions, which are more spread out in comparison with the training values, which are more concentrated. In contrast, the large values in samples with P5 and the apolar beads (C) are mainly due to mean shifts, except for C1-P5, which shows a bimodal profile for the generated molecules. Finally, the combinations of P5 with P4 or P5 give the largest values for both metrics, again because of mean shifts.

We next analysed the chemical functional groups in the generated graphs. First, Figure~\ref{sifig:chem_net_common_fg} shows a chemical network representation of the functional groups (FGs) that are present in the generated molecules for at least 50 bead combinations. The FGs overlap substantially with those of the training dataset. In some cases, the model gives more weight to certain FGs, which nevertheless remain very similar to those in the training set.

\begin{figure}
    \centering
    \includegraphics[width=\linewidth]{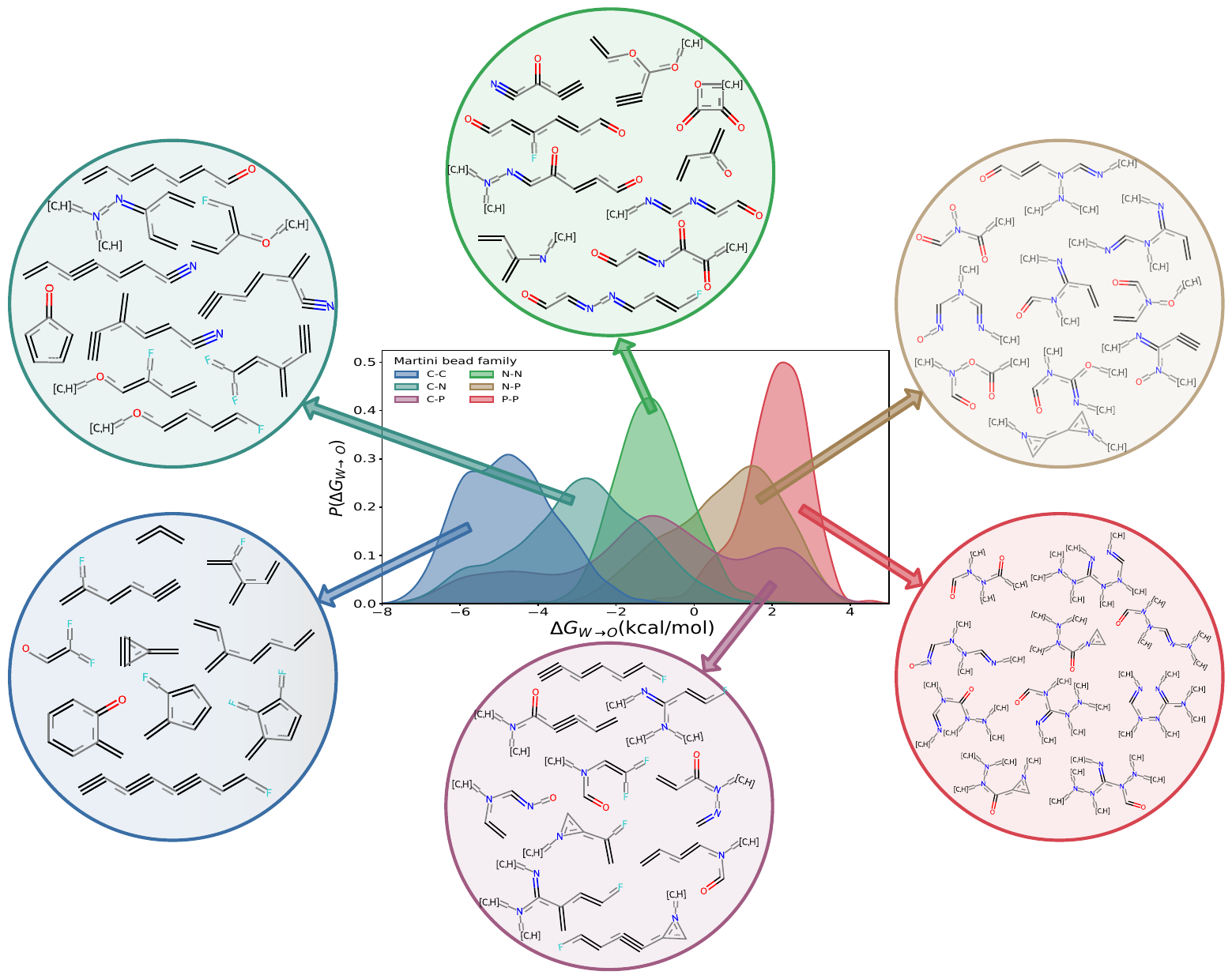}
    \caption{\textbf{Functional groups by bead family.} In the centre, the probability distribution of \dg coloured with respect to the main MARTINI family. The surrounding insets show the ten most common functional-group fragments identified by Ertl's algorithm. Bonds shown in light grey denote query bonds whose order is not fully specified; by default, these match either single or double bonds.}
    \label{fig:fg}
\end{figure}

With the aim of connecting the values of \dg with the presence of FGs, an analysis of the latter by \mar bead family is shown in Figure~\ref{fig:fg}. This figure indicates the presence of specific FGs with respect to the value of \dg that was used as a constraint. Homogeneous combinations of bead classes show the best separation in the \dg distribution. Conversely, cross combinations show values that interpolate between the \dg of the beads that compose them, which is a consequence of considering the values of \dgcg as the sum of individual beads. The figure also shows that the C-P class has the broadest distribution, with three peaks near the mean values of the homogeneous combinations. This agrees with the changes in $\JS$ and \mmd discussed earlier.

From a chemical perspective, the most common FGs identified by Ertl's algorithm show clear trends. Starting with the molecules of the C-C bead family, the most common FGs are long branched hydrocarbons containing multiple carbon--carbon double or triple bonds. Most fragments contain only carbon atoms, with few substitutions by F and a few carbonyls, as could be expected for highly hydrophobic compounds. Moving on to the C-N bead family, the fragments retain the high proportion of hydrocarbons but also show an increase in the number of heteroatoms (N, O), which introduce fragments such as nitriles (\ce{R-C#N}), imines (\ce{R1R2C=N-R3}), aldehydes (\ce{R-CH=O}), and ketones (\ce{R1R2C=O}). Continuing with the N-N bead family, the FGs increase the number of oxygen-containing fragments---ketones, aldehydes, ethers (\ce{R1-O-R2}), and acid anhydrides (\ce{R1C(=O)-O-C(=O)R2})---as well as nitrogen-containing fragments---azo compounds (\ce{R1-N=N-R2}), amides (\ce{R-C(=O)NH2}), and imines---while the number of carbon atoms decreases relative to that of heteroatoms. For the N-P family, the FGs are more complex and polar. In particular, we observe the same fragments as in the other families, as well as imides (\ce{R1C(=O)-NH-C(=O)R2}), isocyanates (\ce{R-N=C=O}), and ethyleneimine rings. The P-P family is characterised by heteroatoms at the core of the FGs, together with all the previously described fragments. Hydrocarbon moieties are minimal and are found only at the termini of the FGs. Multiple \ce{C-C} bonds are absent, replaced by \ce{N-N} or \ce{C-N} bonds. Finally, the C-P family shows a large diversity of FGs: hydrocarbon fragments appear at almost the same rate as fragments containing N, O, and F, and the same holds for the number of multiple \ce{C-C} bonds. Notably, most of the identified FGs are very large, reflecting the large, stable molecules the model generates for this family. In summary, the FG analysis shows that each family has distinct chemistry, with identifiable fragments that map directly onto the values of \dg. It also confirms that the values of \dg and, consequently, the backmapping are, on average, correct, because the generated molecules contain FGs appropriate to the imposed constraint.

Complementary quantitative metrics to analyse the chemical space, described in Section~\ref{subsec:chem_ana}, are presented in Figures~\ref{sifig:icd} and \ref{sifig:fcd}. The internal chemical diversity (ICD) in Figure~\ref{sifig:icd} takes large values, with a minimum of 0.7 and, in most cases, between 0.8 and 0.9. This indicates that the generated samples differ substantially from one another. The C-C family is of particular interest: the intrinsically low diversity of hydrophobic fragments, mainly hydrocarbons, explains the reduced diversity of the generated samples. The ICD can also be examined as a function of $\gamma$. In general, large values of $\gamma$ reduce diversity in the generated samples, although they strengthen the conditioning~\cite{buzzard2025understanding}. Previous applications in the prediction of HOMO--LUMO gaps~\cite{ninniri2024classifier} made use of large values of $\gamma$ with the aim of enforcing the property more strongly. This is not ideal for our work, because we want to preserve a wide diversity of compounds while still enforcing the condition. The results in Figure~\ref{sifig:slope_icd} show, in general, a small but consistent decrease of the average ICD as the value of $\gamma$ increases. The largest ICD decreases occur for bead combinations with low population in the training dataset, while the combinations with the largest population are the least affected. The fraction of valid graphs follows the same pattern: high and nearly uniform across bead combinations at $\gamma = 0.5$ and $\gamma = 1$, it degrades at $\gamma = 2$ towards both ends of the hydrophobicity scale, most visibly for the P3--P5 combinations and to a lesser extent for those built from C1--C3 (Figure~\ref{sifig:basic_metrics}). These are the sparsely populated regions in which the conditioning is hardest to satisfy, and enforcing it more strongly there yields a larger fraction of graphs that are disconnected or cannot be resolved into a molecule by RDKit. For this work, sampling from the conditional distribution provides the best results, as smaller values of $\gamma$ reduce the impact of the condition while larger values cost both diversity and validity.

The other quantitative metric studied in this work is the FCD, shown in Figure~\ref{sifig:fcd}. This metric is equivalent to the $L^{2}$-Wasserstein distance between the distribution of the molecules in the training dataset and that of the generated ones. The results indicate that the generated molecules are consistent with those in the training dataset, with small FCD values. Once more, molecules generated for combinations of the P5 bead show the largest distances. Combinations of members of the intermediate family have intermediate FCD values, while the rest are small. Notably, FCD increases for \dg values with low populations in the training dataset. Although this may seem contradictory, FCD and ICD differ: FCD measures how different the generated molecules are from the training samples, while ICD measures how different they are from one another. This can be understood as the model efficiently sampling within the distribution created for each value of \dg.

\section{Conclusions}

This work introduces a generative machine learning model called \jun that bridges CG representations in the \mar model with AA molecules through a graph diffusion model constrained by the value of \dgcg. By leveraging the relationship between bead assignment and \dg, and by using \dgcg as a proxy for the mapping of chemical space by the \mar force field, the model can generate multiple samples and thereby bypass the major complications of compositional backmapping. The two representations are connected by the hypothesis that \dgcg is approximately equal to the average \dgaa value of molecules mapped to a given bead combination.

The results show that \jun can be efficiently conditioned to generate samples whose \dg follows the value of \dgcg; other physicochemical properties of interest, such as the solvent-accessible surface area, are recovered from the training dataset as well. Analysis of the differences between the mean \dgaa and the constrained \dgcg passed to \jun shows a linear relationship between these quantities, with deviations for the most hydrophobic and hydrophilic samples due to limitations in the \dg predictor. Furthermore, comparing the \dg distributions for training and generated samples shows strong agreement. A study of functional groups, and hence of the chemical content of the generated samples, provides further evidence that the bead combinations translate into AA samples under the conditioning on \dgcg. The fact that the functional groups are separated by bead family, with no previous chemical information given to the model beyond a single free energy, suggests that \dg carries more of the chemistry of the \mar mapping than its role as one fitting target among several would imply.

Several of our results hold more generally for generative chemistry. We analysed the effect of the hyperparameter $\gamma$, which controls the balance between conditioned and unconditioned probabilities for sample generation. Large values of $\gamma$ improve the generation of samples with property values near the desired value, at the cost of the diversity and, in the tails of the distribution, the validity of the generated molecules. The effect of $\gamma$ is also more pronounced there than at the mean. Finally, the choice of estimator for the conditioning property matters, and matters most in the tails: an additive atom-contribution scheme is a poor instrument for a property as collective as partitioning.

The work presented here is limited to small molecules with a maximum of 9 heavy atoms that map onto a two-bead \mar v2 representation. Unimers entered the training set, but the 1,200 samples generated for them---against 38,600 for dimers---are too few to support a dedicated analysis, and the results reported above therefore concern dimers alone. For backmapping compounds with more than two beads, it is necessary to include additional information on the topology of the CG representation. Relatedly, this work does not consider the effect of bonded parameters (i.e., the distance between two beads). We expect that incorporating such information will improve the performance of the backmapping process. A further restriction is that we assess the generated compounds on validity, hydrophobicity, and chemical composition, but not on synthetic accessibility; a generated graph is a well-formed molecule, not necessarily a preparable one, and coupling the conditioning to a retrosynthetic score is a natural extension. We plan to address these aspects in future work.

More broadly, this work is a step toward closing the screening or optimisation loop in chemical compound space using CG models. Using the model developed here, promising bead combinations identified at the CG level can be converted back to AA resolution, so that the generated candidate molecules match the CG-level hydrophobicity and carry chemistries distinguishable by functional-group composition. These candidates can then be inspected or simulated at the AA level. For the dimers studied here, this return step is feasible, so the compression of chemical space that makes low-resolution screening cheap no longer ends at a bead string: that string can now be carried back to real molecules.

\section*{Author contributions}
LIVS and TB conceived the study and designed the methodology. LIVS implemented the
software, performed the training and generation, carried out the formal analysis and
validation, curated the data, produced the figures, and wrote the original draft. TB
provided supervision and computational resources. Both authors acquired funding, and
reviewed and edited the manuscript.

\section*{Conflicts of interest}
There are no conflicts to declare.

\section*{Data availability}
An interactive deployment of \jun is available at \url{https://livazquezs-juniper-dg.hf.space/}. The source code and the scripts used to generate molecules can be found at \url{https://github.com/LIVazquezS/juniper_dg}. The trained model
weights can be found at \url{https://doi.org/10.5281/zenodo.22282325}. Scripts to analyse the results are available at \url{https://github.com/LIVazquezS/Juniper_analysis}. The unimer
and dimer training sets are those of Ref.~\citenum{menichetti2019drug} and are not
redistributed here. The generated molecules underlying the analyses reported here are deposited
at \url{https://doi.org/10.5281/zenodo.22259586}. Supplementary figures, tables, and
further methodological detail are provided in the electronic supplementary information
(ESI).

\section*{Acknowledgements}
The authors thank Sander Hummerich and Luis Walter for critical reading and for their valuable feedback on
the manuscript. LIVS acknowledges funding from the Swiss National Science
Foundation (Grant P500PN\_222297). The authors acknowledge support by the state of Baden-W\"urttemberg through bwHPC
and the German Research Foundation (DFG) through grant INST 35/1597-1 FUGG.

\section*{Declaration of AI-assisted tools}
 
During the preparation of this manuscript, the authors used Claude (Anthropic;
Opus 4.8 and Opus 5, accessed August 2026) for language editing, structural revision of the text, and
\LaTeX{} preparation of the manuscript and electronic supplementary information. All data, analyses,
and figures are the authors' own. The authors reviewed, edited, and approved all text and take full
responsibility for the content.

\clearpage
\bibliographystyle{achemso}
\bibliography{refs}
\clearpage
\appendix

\renewcommand{\thepage}{S\arabic{page}}
\renewcommand{\thetable}{S\arabic{table}}
\renewcommand{\thefigure}{S\arabic{figure}}
\renewcommand{\theequation}{S\arabic{equation}}

\setcounter{page}{1}
\setcounter{figure}{0}

\section{Discrete graph diffusion: full formulation}
\label{sisec:diffusion}

This section states in full the discrete denoising diffusion construction summarised in Section~\ref{subsec:diffusion} of the main manuscript. The construction follows the \textsc{DiGress} model of Vignac \textit{et al.}~\cite{vignac2022digress} together with the classifier-free conditioning of Ninniri \textit{et al.}~\cite{ninniri2024classifier}. It is reproduced here for completeness and contains no development original to this work.

\subsection{Forward process (noising)}

As mentioned before, molecules are considered graphs $G = (\mathcal{X},\mathcal{E},y)$, where the space $\mathcal{X}$ contains $x_{i}$ features for each node $i$ with a one-hot encoding for each atom, $x_{i}\in \mathbb{R}^{a}$, and $a$ is the cardinality of the space $\mathcal{X}$ (i.e., the number of atom types). For edges, each bond type is also one-hot encoded as $e_{ij} \in \mathbb{R}^{b}$, where $b$ is the cardinality of the space $\mathcal{E}$. Additionally, the absence of a bond is explicitly encoded with a special type. The encodings are organised in a matrix $\mathbf{X}\in \mathbb{R}^{n \times a}$ for nodes and $\mathbf{E}\in \mathbb{R}^{n\times n \times b}$ for edges. In both cases, $n$ is the number of nodes. Given that the space of a complete graph is too large to be handled at once, we treat edges and nodes separately. Therefore, we diffuse separately over the nodes' and edges' features.

For the diffusion process, we define transition matrices over nodes and edges as:

\begin{equation*}
    [\mathbf{Q}^{t}_{X}]_{ij} = q(x^{t}=j \mid x^{t-1}=i), \qquad [\mathbf{Q}^{t}_{E}]_{ij} = q(e^{t}=j \mid e^{t-1}=i).
\end{equation*}

Then, the process of adding noise to the graph $G^{t}=(\mathbf{X}^{t},\mathbf{E}^{t},y)$ is governed by:

\begin{equation}
    q(G^{t} \mid G^{t-1}) = (\mathbf{X}^{t-1}\mathbf{Q}^{t}_{X}, \mathbf{E}^{t-1}\mathbf{Q}^{t}_{E}).
    \label{eq:noising}
\end{equation}

Here $t \sim \mathcal{U}(1,\dots,T)$, $\mathbf{X}^{0} = \mathcal{X}$, and $\mathbf{E}^{0} = \mathcal{E}$. The process in Eq.~\ref{eq:noising} is generalised over $T$ steps as:

\begin{equation*}
    q(G^{t} \mid G) = (\mathbf{X}\overline{\mathbf{Q}}^{t}_{X}, \mathbf{E}\overline{\mathbf{Q}}^{t}_{E}),
\end{equation*}

where the cumulative transition matrices for nodes or edges are:

\begin{equation*}
    \overline{\mathbf{Q}}^{t}_{X/E} = \prod_{\tau=1}^{t} \mathbf{Q}^{\tau}_{X/E}.
\end{equation*}

The transition matrices of edges and nodes can be chosen arbitrarily. However, we impose the condition

\begin{equation*}
    \lim_{T\rightarrow\infty} \mathbf{Q}^{T}_{X}\mathds{1}_{i} = \mathbf{m}_{X} \quad \forall i,
\end{equation*}

where $\mathbf{m}_{X}$ is the marginal distribution of nodes (i.e., atoms). A similar expression can be obtained for edges. Following this condition, the proposed transition matrices are:
\begin{equation}
    \mathbf{Q}^{t}_{X/E} = \alpha^{t} \mathbf{I} + \beta^{t}\mathbf{1}\mathbf{m}_{X/E}.
    \label{eq:noise}
\end{equation}

In this case, the probability of a transition from state $i$ to state $j$ is proportional to the marginal probability of category $j$ in the training dataset. Eq.~\ref{eq:noise} is generalised as:

\begin{equation*}
    \overline{\mathbf{Q}}^{t} = \overline{\alpha}^{t} \mathbf{I} + \overline{\beta}^{t}\mathbf{1}\mathbf{m},
\end{equation*}
where $\overline{\alpha}^{t} = \prod_{\tau=1}^{t} \alpha^{\tau}$, defined by the cosine schedule $\overline{\alpha}^{t} = \cos^{2}\!\left[\frac{\pi}{2}\,\frac{t/T+s}{1+s}\right]$ with $s\ll 1$. Correspondingly, $\beta^{t} = 1-\alpha^{t}$.

\subsection{Reverse process (inference)}

In the denoising process, a neural network (NN) $\Phi_{\theta}$ parametrised by $\theta$ takes as input $G^{t}$, a noisy graph, to predict a clean graph ($G = \Phi_{\theta}(G^{t})$).

The NN model is trained to estimate the probability of the reverse diffusion iteration, $P_{\theta}(G^{t-1} \mid G^{t})$. Here, these probabilities are conditioned on the value of the global label $y$ (\dgcg) corresponding to each graph, $P_{\theta}(G^{t-1} \mid G^{t},y)$. The value of $y$ remains constant throughout the noising process. There are different ways of computing the desired probabilities; here, the classifier-free denoising procedure introduced by Ninniri \textit{et al.}~\cite{ninniri2024classifier} was used.

In the classifier-free model, the conditional probability for denoising a graph conditioned on the value of $y$ is given by:

\begin{equation}
    P_{\theta}(G^{t-1} \mid G^{t},y) = \prod_{1\leq i \leq n} p_{\theta} (\mathbf{x}_{i}^{t-1} \mid G^{t},y)\cdot \prod_{1\leq i\leq n} \prod_{1 \leq j \leq n} p_{\theta}(\mathbf{e}_{ij}^{t-1} \mid G^{t},y).
\end{equation}

This result is marginalised over the predictions of the NN model and uses the result of Dhariwal and Nichol~\cite{dhariwal2021diffusion} on the separation of conditional probabilities. Using these results, the marginal probabilities for nodes are given by

\begin{equation}
    p_{\theta}(\mathbf{x}_{i}^{t-1} \mid G^{t},y) = \underbrace{\sum_{x\in \mathcal{X}} q(\mathbf{x}_{i}^{t-1} \mid \mathbf{x}_{i}^{t},\mathbf{x}_{i}^{0}=x)}_{\text{transition probability}}\cdot \underbrace{\Phi_{\theta}(\mathbf{x}_{i}^{0}=x \mid G^{t}, y)}_{\text{neural network}}.
    \label{eq:mprob_nodes}
\end{equation}

Similarly, for the edges:
\begin{equation}
    p_{\theta}(\mathbf{e}_{ij}^{t-1} \mid G^{t},y) = \sum_{e\in \mathcal{E}} q(\mathbf{e}_{ij}^{t-1} \mid \mathbf{e}_{ij}^{t},\mathbf{e}_{ij}^{0}=e)\cdot \Phi_{\theta}(\mathbf{e}_{ij}^{0}=e \mid G^{t}, y).
    \label{eq:mprob_edges}
\end{equation}

In Eqs.~\ref{eq:mprob_nodes} and \ref{eq:mprob_edges}, the first term on the right-hand side is the transition probability matrix, while the second is the NN model that predicts the node/edge types conditioned on a noisy graph ($G^{t}$) and the guide value ($y$).

These marginals enter the barycentric combination of conditional and unconditional predictions, Eq.~\ref{eq:cond} of the main manuscript, from which samples are drawn.

\section{Conditional variational autoencoder}

To benchmark our approach against other property-conditioned generative models, we ported the model of Ref.~\citenum{lim2018molecular} to PyTorch. This model is a Conditional Variational Autoencoder (CVAE) built on recurrent neural networks, using an LSTM cell in both the encoder and the decoder. We adopted the same hyperparameters as reported in Ref.~\citenum{lim2018molecular}. For an initial validation, we trained the model on the dataset provided in the original repository, conditioning on 3 properties rather than the 5 used in the original paper. The generated molecules show a high correlation with the target property values, which we take as evidence of adequate performance.

We introduced several modifications to reduce overfitting and improve performance. First, we applied a dropout of 0.2 to the LSTM cell. We also modified the optimiser by adding a weight decay of $10^{-5}$ and a decay rate of 0.97. Finally, we implemented early stopping with a patience of 10 epochs.

We trained two models on the dataset used in this work. The first, referred to as ``CVAE Mult. Prop.'', was conditioned on 5 properties: molecular weight, \dgcg, hydrogen-bond donors, hydrogen-bond acceptors, and topological polar surface area (TPSA). The second, referred to as ``CVAE Sing. Prop.'', was conditioned only on \dgcg. Both models were trained for a maximum of 200 epochs but stopped at epoch 115, as the validation loss had ceased to improve. The model is archived at \url{https://github.com/LIVazquezS/CVAE}.

\section{Why juniper?}

Martini cocktails are made by mixing gin and vermouth. In the case of the coarse-grained force field, the ``gin'' can be considered to be the beads and their types, while the ``vermouth'' is the mapping. Because we aim to recover the components of the ``gin,'' we call our method \jun, since juniper berries are the main flavouring of gin.

\clearpage

\section{Tables}

\begin{table}[h]
\small
\setlength{\tabcolsep}{4pt}
\begin{tabular}{c|ccccc}
Model                   & Validity & Uniqueness & Novelty & Repetition & Coverage  \\ \hline
JT-VAE* (ZINC-250K)      & 1.00     & 1.00   & 0.89  & 0.00       & 0.00  \\
DiGress* (ZINC-250K)     & 0.85     & 1.00   & 1.00  & 0.15       & 0.00  \\
MG-DIFF* (ZINC-250K)     & 0.96     & 0.99   & 0.99  & 0.03       & 0.00  \\
DrugDiff $\dagger$ (ZINC-250K)    & 1.00     & 0.99   & 1.00  & 0.01       & 0.00  \\ \hline
CVAE Mult. Prop. (CG dataset)         & 0.71     & 0.71   & 0.50  & 0.00       & 0.21  \\
CVAE Sing. Prop. (CG dataset)        & 0.73     & 0.72   & 0.47  & 0.00       & 0.25  \\
Juniper no guidance (CG dataset)     & 0.95     & 0.94   & 0.50  & 0.00       & 0.45  \\
Juniper dimer guidance (CG dataset) & 0.93     & 0.92   & 0.44  & 0.00       & 0.48
\end{tabular}
\caption{Performance metrics for state-of-the-art models for the generation of molecular graphs. Data with an * is taken from Ref.~\citenum{zhang2025mg} and $\dagger$ is taken from Ref.~\citenum{oestreich2025drugdiff}. The dataset used to train the models is shown in brackets. Validity, uniqueness, and novelty are reported as fractions; the repetition rate and coverage follow Eqs.~\ref{eq:repeat} and \ref{eq:coverage} of the main manuscript, and all five metrics are defined in Section~\ref{subsec:basic_met} there.}
\label{sitable:perfom_other}
\end{table}

\begin{table}[h]
\small
\setlength{\tabcolsep}{4pt}
\begin{tabular}{cccc}
\multicolumn{1}{c|}{Property} & Train/Juniper            & Train/CVAE Single        & Train/CVAE Mult. Prop. \\ \hline
\multicolumn{1}{c|}{\dg}     & 0.004                    & 0.012                    & 0.007                  \\
\multicolumn{1}{c|}{SASA}     & 0.005                    & 0.013                    & 0.005                  \\
                              &                          &                          &                        \\ \hline \hline
& Juniper/CVAE Sing. Prop. & Juniper/CVAE Mult. Prop. &                        \\\hline
\multicolumn{1}{c|}{\dg}     & 0.011                    & 0.007                    &                        \\
\multicolumn{1}{c|}{SASA}     & 0.011                    & 0.007                    &                       
\end{tabular}
\caption{Jensen--Shannon divergence between the distributions of \dg and SASA for the generated samples without guidance by Juniper and CVAE models. In the first half, we calculate $D_{\mathrm{JS}}$ between the training and unconditioned distributions. The second half shows values for comparison between the models' distributions.}
\label{sitable:JS_others}
\end{table}

\newpage

\section{Figures}

\begin{figure}[h]
    \centering
    \includegraphics[width=0.9\linewidth]{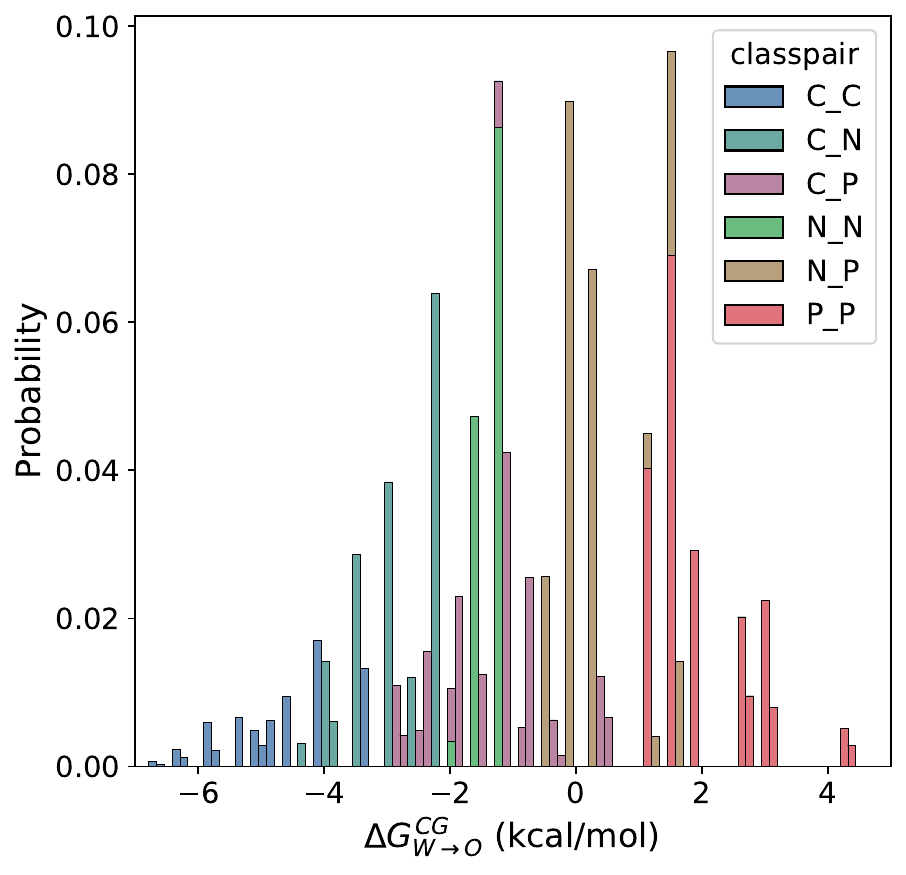}
    \caption{Histogram of bead population as a function of \dg for samples in the training dataset.}
    \label{sifig:train_hist}
\end{figure}

\begin{figure}
    \centering
    \includegraphics[width=\linewidth]{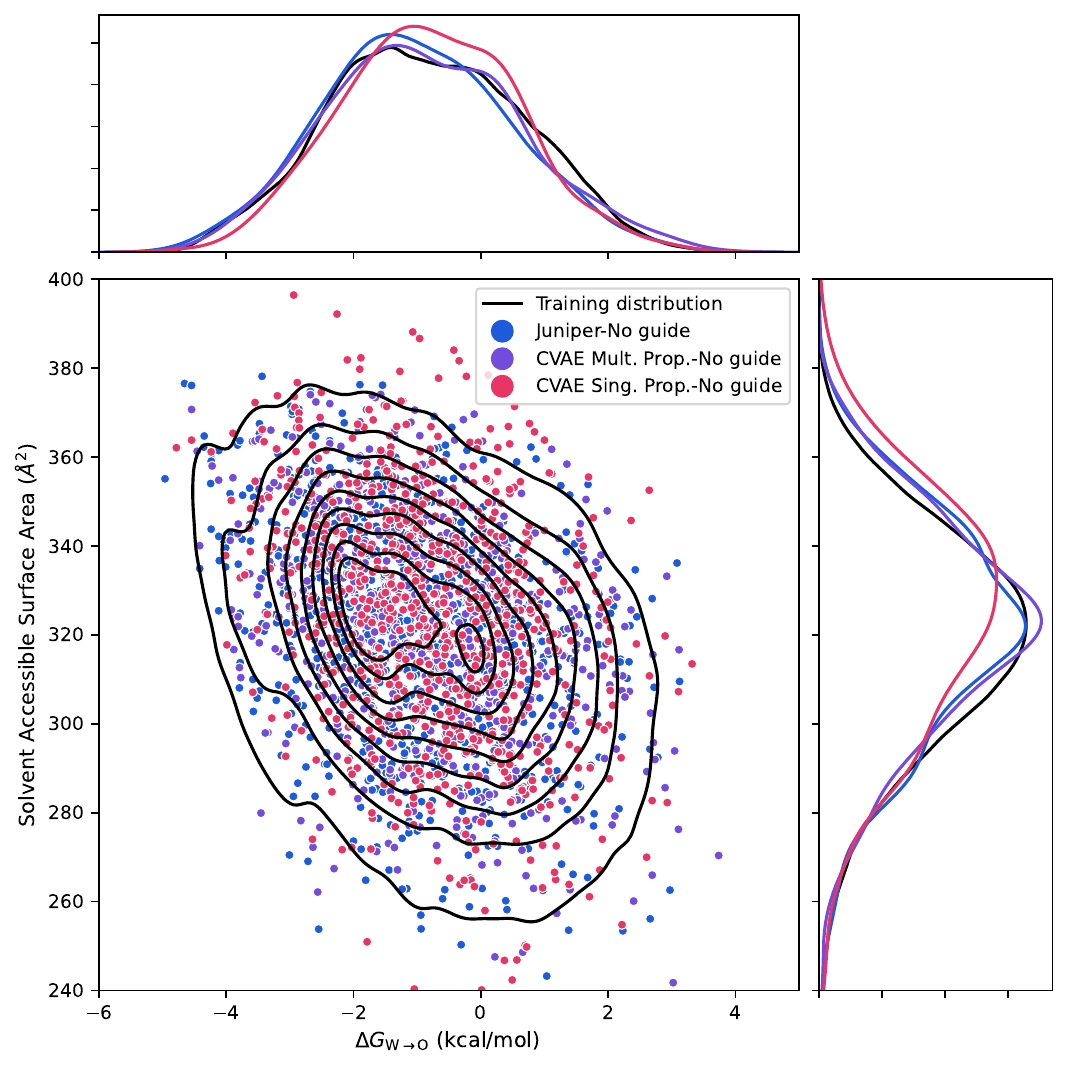}
    \caption{\textbf{Unguided sampling.} 2D kernel density representation of the chemical space in terms of the octanol--water partition free energy and the solvent-accessible surface area of the training dataset (black line). The samples generated without explicit guidance are shown in different colours by the model used to generate them. The figure margins show 1D kernel density estimates for each quantity.}
    \label{sifig:unguid_samp}
\end{figure}

\begin{figure}
    \centering
    \includegraphics[width=\linewidth]{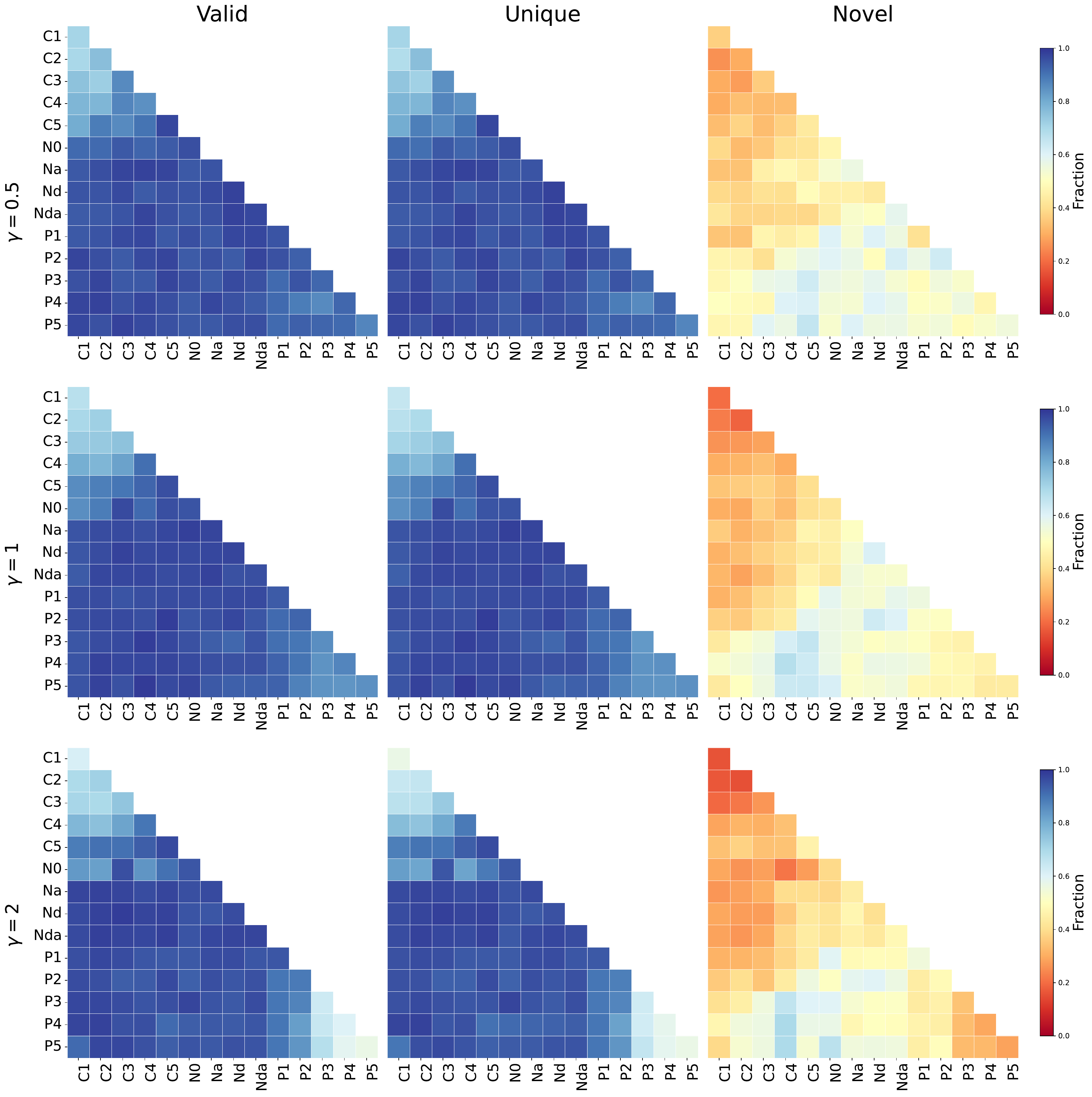}
    \caption{Basic metrics for the generated molecules for dimers. Left: fraction of valid molecules. Centre: fraction of unique generated samples. Right: fraction of novel molecules among those that are valid and unique. From top to bottom, the value of $\gamma$ increases.}
    \label{sifig:basic_metrics}
\end{figure}

\begin{figure}
    \centering
    \includegraphics[width=0.8\linewidth]{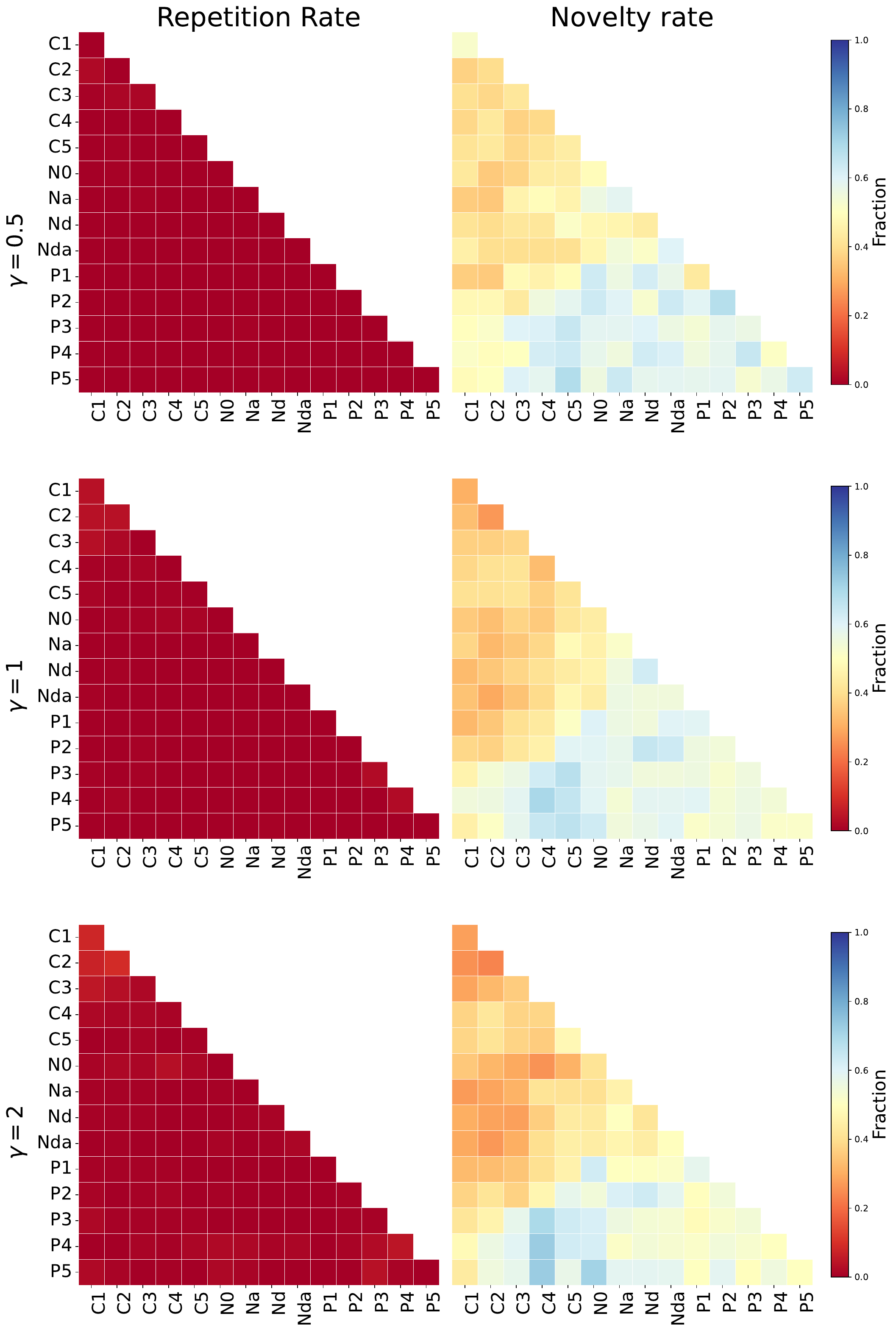}
    \caption{Repetition and novelty rate for dimers. On the left, the repetition rate as defined in Eq.~\ref{eq:repeat} of the main manuscript. On the right, the novelty rate from Eq.~\ref{eq:novel}. From top to bottom, the value of $\gamma$ increases.}
    \label{sifig:rep_nov_rate}
\end{figure}

\begin{figure}
    \centering
    \includegraphics[width=\linewidth]{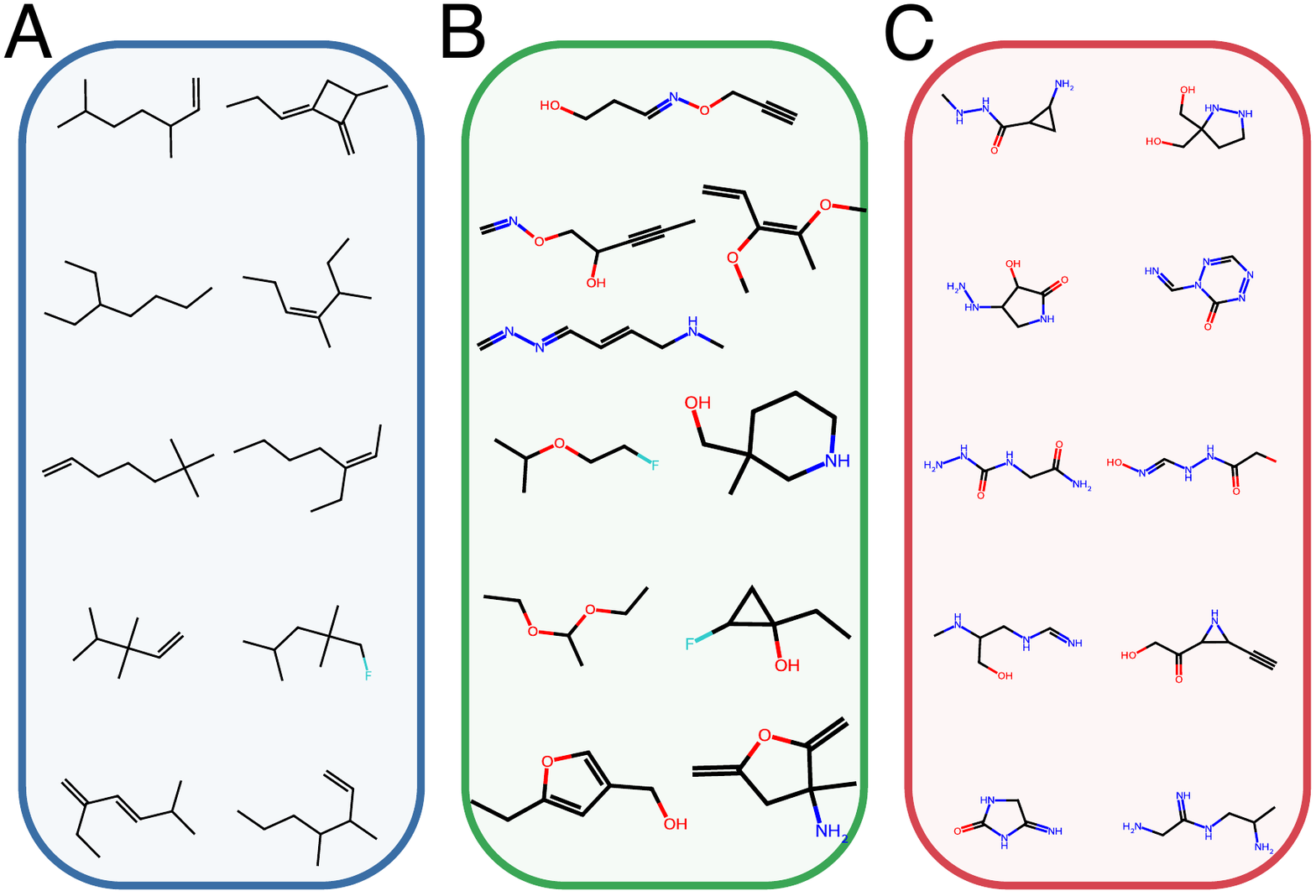}
    \caption{Examples of molecules generated at the specific values of \dg used in Figure~\ref{fig:2d} of the main manuscript. From left to right: (A) C1-C1, (B) Na-Nda, and (C) P3-P4.}
    \label{sifig:mols_examples}
\end{figure}

\begin{figure}
    \centering
    \includegraphics[width=\linewidth]{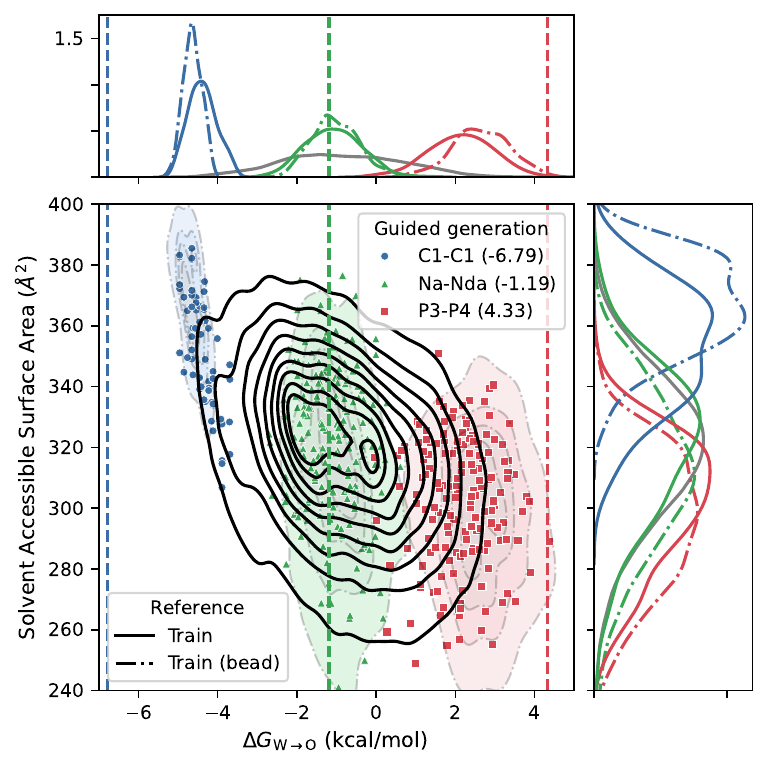}
    \caption{\textbf{Sampling in chemical space.} 2D kernel density estimate of the joint distribution of \dg and the solvent-accessible surface area (SASA). The figure margins show 1D kernel density estimates for each quantity. In the background, the solid black line shows the distribution for all molecules in the training dataset. The scattered points represent the conditional samples generated for three cases: C1-C1 (very hydrophobic), Na-Nda (intermediate), and P3-P4 (hydrophilic). Distributions of each property by bead of the training dataset are shown as colour-filled representations and with dashed-dotted lines in the figure margins. The effect of the conditioning on the sampling is visible in all three cases. Values of \dg were obtained with RDKit. Figure~\ref{sifig:mols_examples} shows examples of the molecules generated for the \dg value corresponding to each bead combination.}
    \label{sifig:2d_rdkit}
\end{figure}

\begin{figure}
    \centering
    \includegraphics[width=0.7\linewidth]{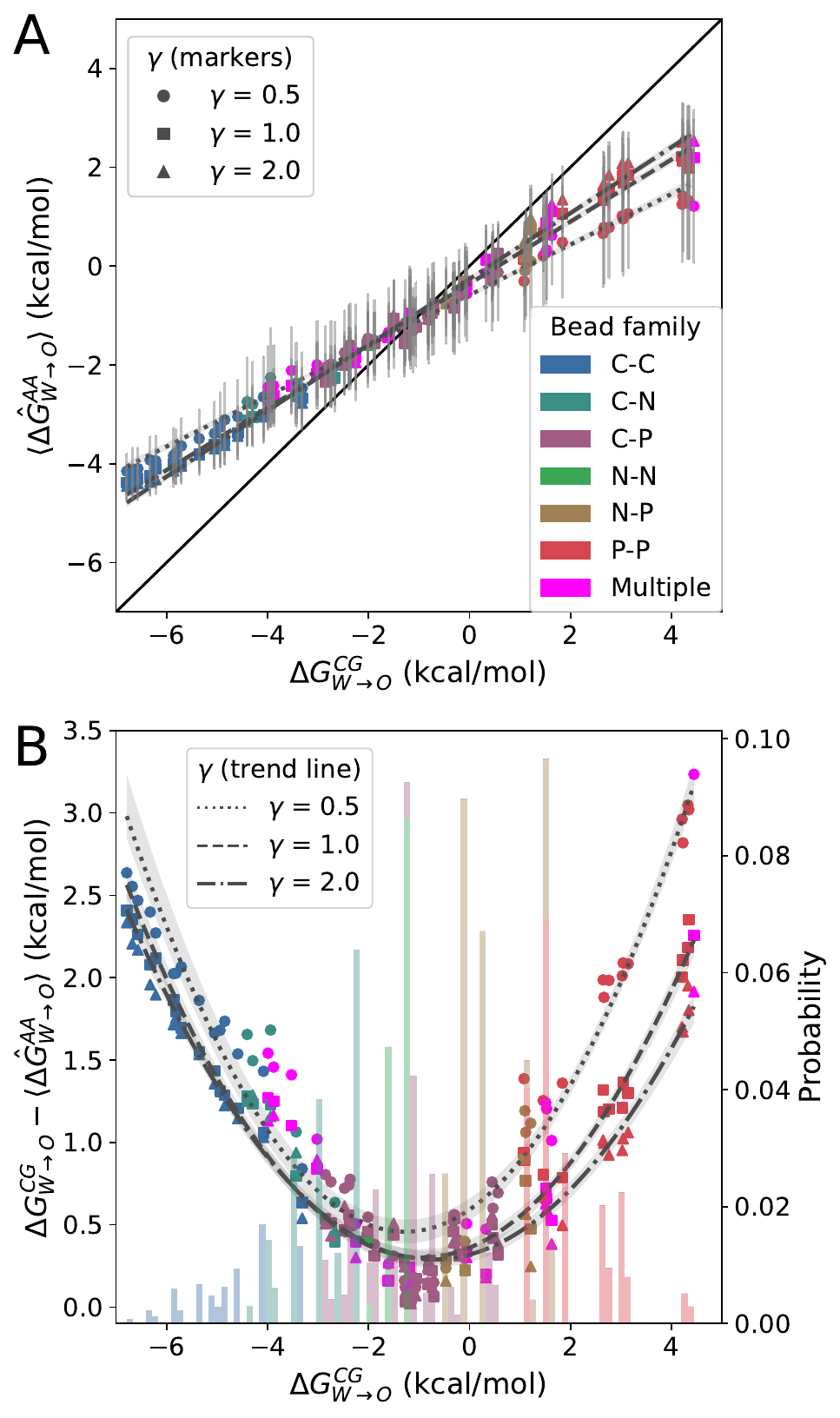}
    \caption{\textbf{Comparison between \dgcg and \dgaa.} (A) Scatter plot of the value of \dgcg versus the mean value of \dgaa for the generated molecules. The markers are coloured by bead family, except for values of \dg that can correspond to multiple bead combinations, which are coloured in magenta. (B) Residual plot of the absolute difference between the mean of \dgaa and the value of \dgcg for all combinations. A trend line fitted to a second-order polynomial is shown for each value of $\gamma$. The background shows the histogram of the training dataset.}
    \label{sifig:diff_dg_aacg_rdkit}
\end{figure}

\begin{figure}
    \centering
    \includegraphics[width=0.5\linewidth]{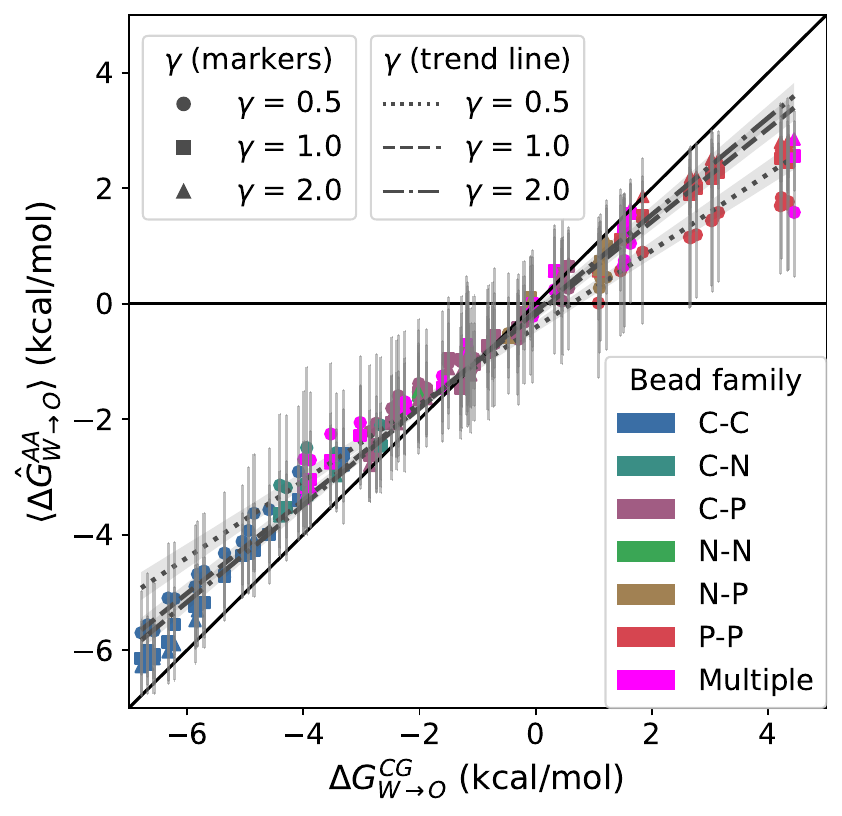}
    \caption{Signed residual plot for the difference between \dgcg and the mean value of \dgaa using ALOGPS.}
    \label{sifig:signed_dgaacg}
\end{figure}

\begin{figure}
    \centering
    \includegraphics[width=\linewidth]{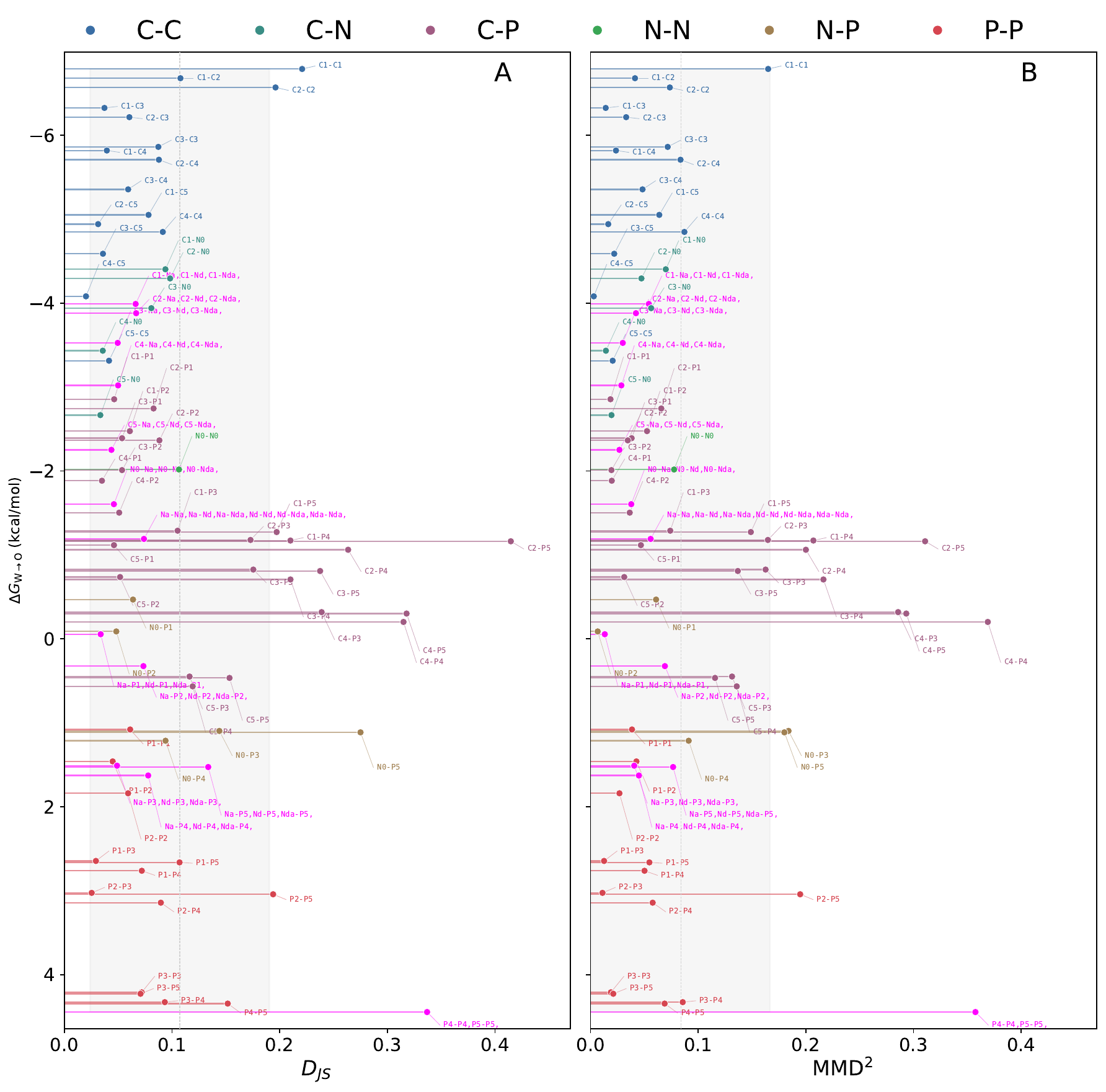}
    \caption{Statistical distances between the distributions of \dg of training and generated samples with a value of $\gamma=1$. (A) Jensen--Shannon divergence (cf.\ Eq.~\ref{eq:js} of the main manuscript). (B) Square of the maximum mean discrepancy (cf.\ Eq.~\ref{eq:mmd}). The dotted green line shows the average value, while the shaded region corresponds to $\pm$ one standard deviation. For the MMD, plots of the generated and reference distributions as well as the witness function for each combination are shown in Figure~\ref{sifig:witness_s1}.}
    \label{sifig:stat_dist}
\end{figure}

\begin{figure}
    \centering
    \includegraphics[width=\textwidth]{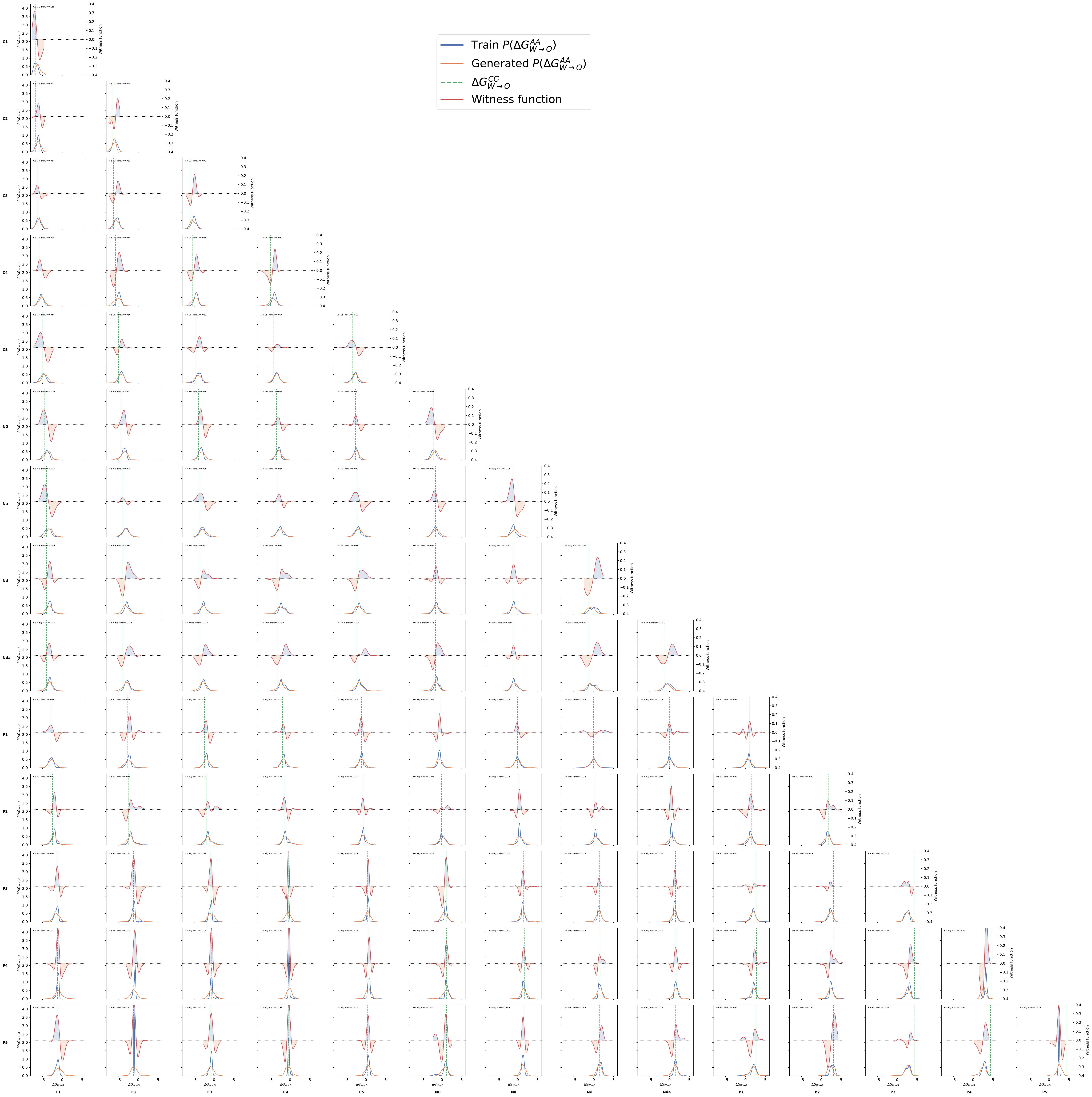}
    \caption{Distributions of $P(\Delta G^{\mathrm{train}}_{\mathrm{W} \mapsto \mathrm{O}})$ and $P(\Delta G^{\mathrm{gen}}_{\mathrm{W} \mapsto \mathrm{O}})$, and witness function, for dimer combinations using $\gamma=1$.}
    \label{sifig:witness_s1}
\end{figure}

 \begin{figure}
    \centering
     \includegraphics[width=\linewidth]{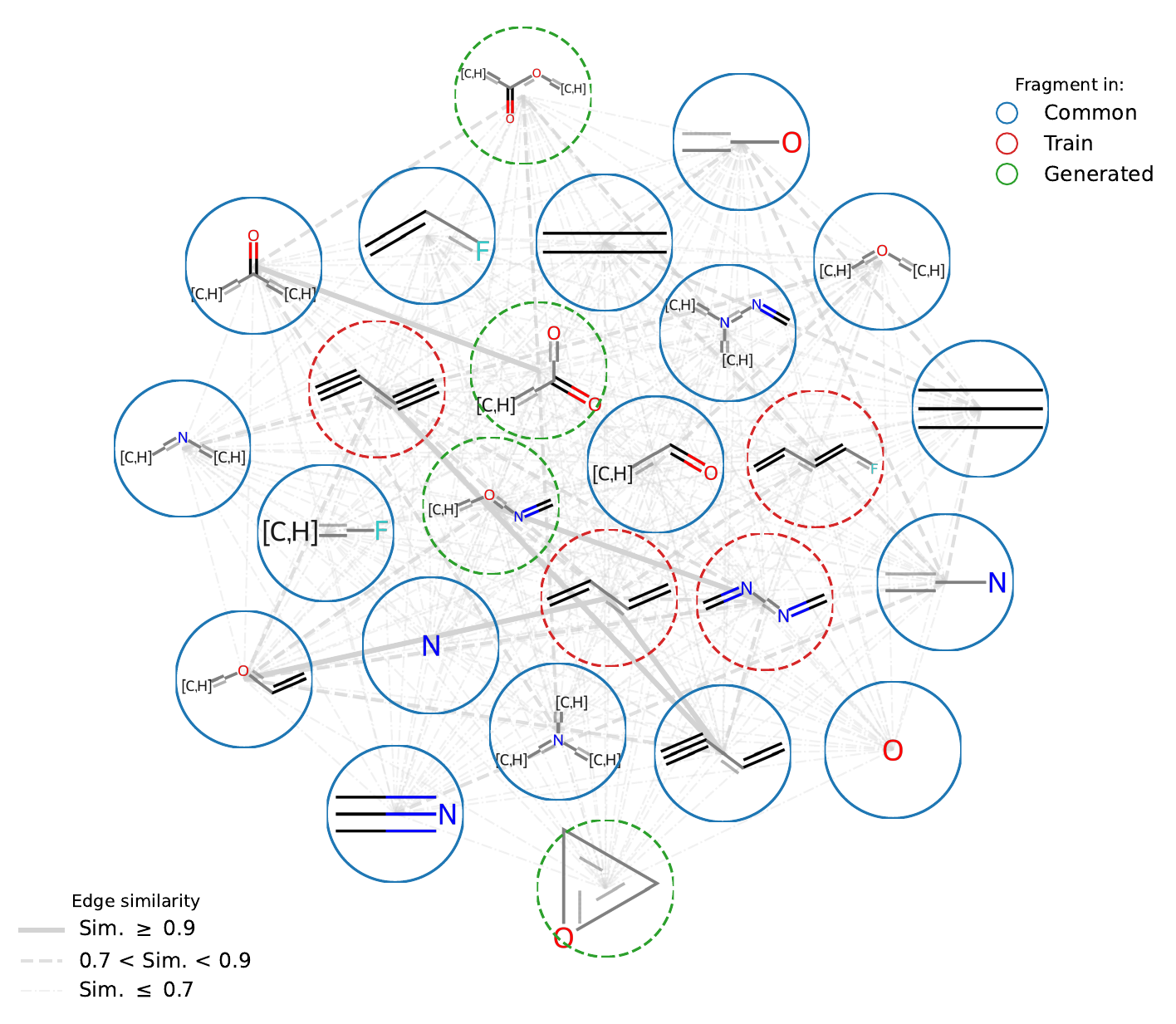}
     \caption{Chemical network comparison of common functional groups between the training and generated sets. The functional groups shown are present in at least 50 bead combinations. The network structure was determined using the t-SNE decomposition of the SMARTS fragments obtained from Ertl's functional group decomposition, ranked by the similarity between fragments. The FGs shared between the training and generated sets are circled in blue, FGs present only in the training set are circled in red, and FGs present only in the generated set are highlighted in green.}
     \label{sifig:chem_net_common_fg}
\end{figure}

\begin{figure}
    \centering
    \includegraphics[width=\linewidth]{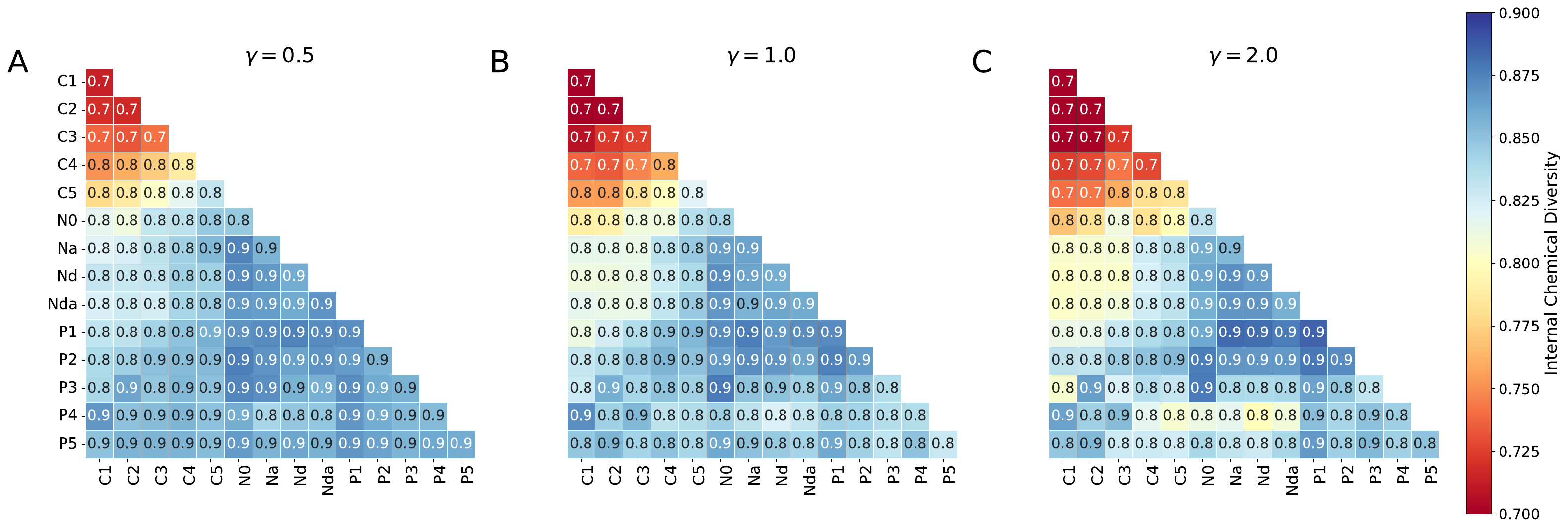}
    \caption{Internal chemical diversity (ICD) on the generated molecules for different combinations of beads. From left to right, the value of $\gamma$ is increased. The colour scale ranges from 0.7 to 0.9. An ICD value closer to one indicates greater diversity in the generated molecules.}
    \label{sifig:icd}
\end{figure}

\begin{figure}
    \centering
    \includegraphics[width=\linewidth]{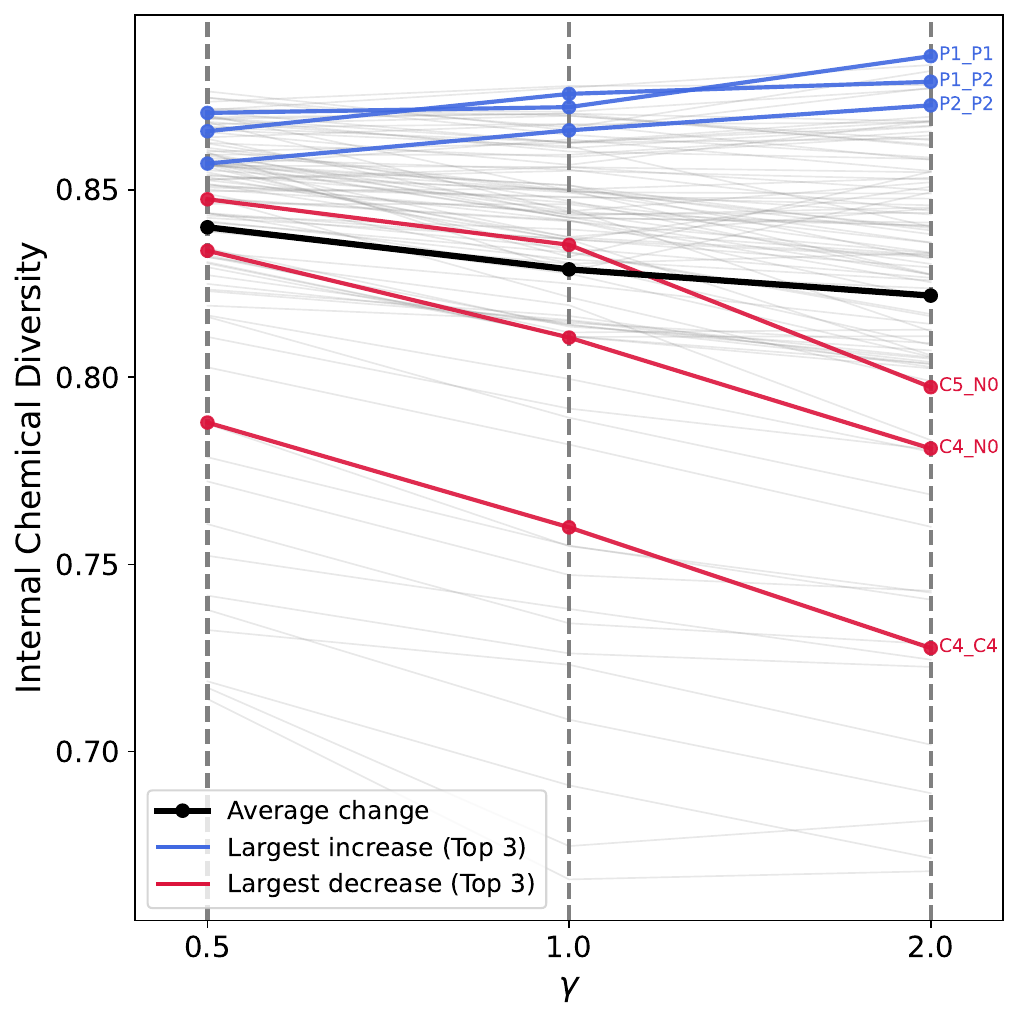}
    \caption{Changes in the value of the internal chemical diversity (ICD) of the generated molecules with respect to the value of $\gamma$. In red, the three bead combinations with the largest decrease in ICD are highlighted. Conversely, the combinations with the largest increase in ICD are coloured in blue. The bold black line shows the mean change across all combinations.}
    \label{sifig:slope_icd}
\end{figure}

\begin{figure}
    \centering
    \includegraphics[width=0.5\linewidth]{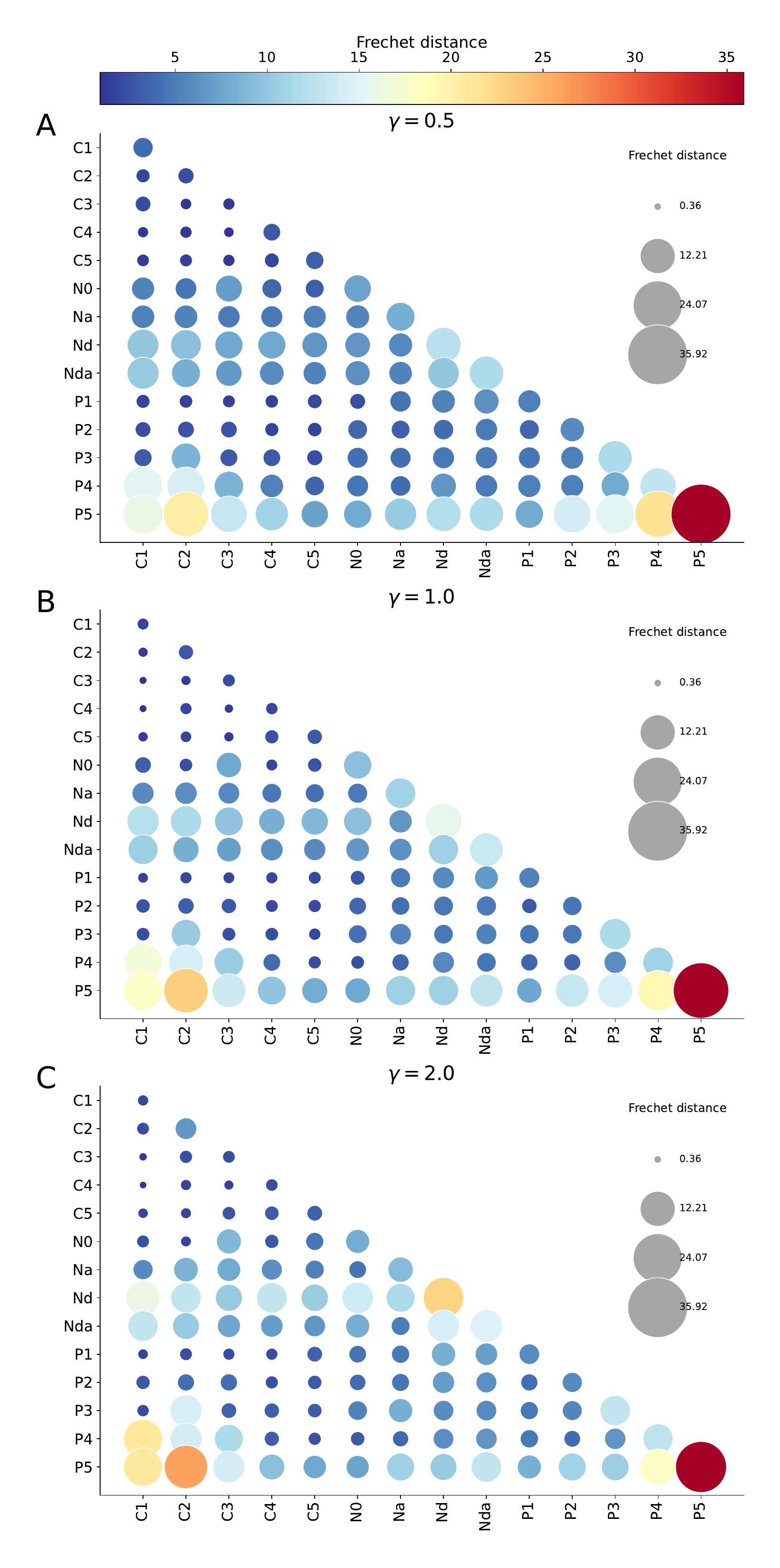}
    \caption{Fr\'echet ChemNet distance (FCD) between the training and generated molecules. The colour code and point radii illustrate the distance between training and generated molecules. From top to bottom, the value of $\gamma$ is increased. Larger values of FCD correspond to larger differences between the structures of the generated compounds and those in the training set.}
    \label{sifig:fcd}
\end{figure}

\end{document}